\documentclass[longauth]{aa}
\usepackage{graphicx}
\usepackage[]{natbib}
\usepackage{multirow}
\usepackage{amsmath} 
\usepackage{txfonts}  
\usepackage{enumitem}

\def\deg{$^{\circ}$}
\def\degb{^{\circ}}

\newcommand{\mum}{\,\mu\hbox{m}}

\newcommand{\Lsun}{L_\odot}

\begin{document}

\title{Two cold belts in the debris disk around the G-type star NZ Lup \thanks{Based on data collected at the European Southern Observatory, Chile under programs 097.C-0523, 097.C-0865, 198.C-0209.}
}

\author{        
A. Boccaletti\inst{\ref{lesia}}
\and P. Th{\'e}bault\inst{\ref{lesia}}
\and N. Pawellek\inst{\ref{mpia}}
\and A.-M. Lagrange\inst{\ref{ipag}}
\and R. Galicher\inst{\ref{lesia}}
\and S. Desidera\inst{\ref{inaf}} 
\and J. Milli\inst{\ref{ipag},\ref{eso}}
\and  Q. Kral\inst{\ref{lesia}}
\and M. Bonnefoy\inst{\ref{ipag}}
\and J.-C. Augereau\inst{\ref{ipag}}
\and A.L. Maire\inst{\ref{mpia},\ref{liege}}
\and T. Henning\inst{\ref{mpia}}
\and H. Beust\inst{\ref{ipag}}
\and L. Rodet\inst{\ref{ipag}}
\and H. Avenhaus\inst{\ref{eth}}  
\and T. Bhowmik\inst{\ref{lesia}} 
\and M. Bonavita\inst{\ref{inaf},\ref{edin}}   
\and G. Chauvin\inst{\ref{ipag}}
\and A. Cheetham\inst{\ref{geneve}}
\and M. Cudel\inst{\ref{ipag}}
\and M. Feldt\inst{\ref{mpia}} 
\and R. Gratton\inst{\ref{inaf}} 
\and J. Hagelberg\inst{\ref{ipag}} 
\and P. Janin-Potiron\inst{\ref{lam}}
\and M. Langlois\inst{\ref{lam},\ref{cral}} 
\and F. M{\'e}nard\inst{\ref{ipag}}
\and D. Mesa\inst{\ref{inaf}} 
\and M. Meyer\inst{\ref{eth},\ref{michigan}}
\and S. Peretti\inst{\ref{geneve}}
\and C. Perrot\inst{\ref{lesia},\ref{valpa},\ref{nucleo}} 
\and T. Schmidt\inst{\ref{lesia}}
\and E. Sissa\inst{\ref{inaf}}  
\and A. Vigan\inst{\ref{lam}} 
\and E. Rickman\inst{\ref{geneve}} 
\and Y. Magnard\inst{\ref{ipag}}
\and D. Maurel\inst{\ref{ipag}}
\and O. Moeller-Nilsson\inst{\ref{mpia}}
\and D. Perret\inst{\ref{lesia}}
\and J.-F. Sauvage\inst{\ref{onera}}
}
 
\institute{
LESIA, Observatoire de Paris, Universit{\'e} PSL, CNRS, Sorbonne Universit{\'e}, Univ. Paris Diderot, Sorbonne Paris Cit{\'e}, 5 place Jules Janssen, 92195 Meudon, France\label{lesia}
\and Univ. Grenoble Alpes, CNRS, IPAG, F-38000 Grenoble, France\label{ipag}
\and Max-Planck-Institut f{\"u}r Astronomie, K{\"o}nigstuhl 17, D-69117 Heidelberg, Germany\label{mpia}
\and INAF-Osservatorio Astronomico di Padova, Vicolo dell’Osservatorio 5, I-35122 Padova, Italy\label{inaf}
\and European Southern Observatory, Alonso de C{\'o}rdova 3107, Casilla 19001 Vitacura, Santiago 19, Chile\label{eso}
\and STAR Institute, Universit{\'e} de Li{\`e}ge, All{\'e}e du Six Ao{\^u}t 19c, B-4000 Li{\`e}ge, Belgium\label{liege}
\and ETH Zurich, Institute for Astronomy, Wolfgang-Pauli-Str. 27, CH- 8093, Zurich, Switzerland\label{eth}
\and SUPA, Institute for Astronomy, The University of Edinburgh, Royal Observatory, Blackford Hill, Edinburgh, EH9 3HJ, UK\label{edin}
\and Geneva Observatory, University of Geneva, Chemin des Mailettes 51, 1290 Versoix, Switzerland\label{geneve}
\and Aix Marseille Universit{\'e}, CNRS, LAM (Laboratoire d’Astrophysique de Marseille) UMR 7326, 13388 Marseille, France\label{lam}
\and CRAL, UMR 5574, CNRS, Universit{\'e} de Lyon, Ecole Normale Sup{\'e}rieure de Lyon, 46 All{\'e}e d’Italie, F–69364 Lyon Cedex 07, France\label{cral}
\and Department of Astronomy, University of Michigan, 1085 S. University, Ann Arbor, MI 48109\label{michigan}
\and Instituto de F{\'i}sica y Astronom{\'i}a, Facultad de Ciencias, Universidad de Valpara{\'i}so, Av. Gran Breta{\~n}a 1111, Valpara{\'i}so, Chile\label{valpa}
\and N\'ucleo Milenio Formaci\'on Planetaria - NPF, Universidad de Valpara\'iso, Av. Gran Breta\~na 1111, Valpara\'iso, Chile  \label{nucleo}
\and DOTA, ONERA, Universit{\'e} Paris Saclay, F-91123, Palaiseau France\label{onera}
}

 \offprints{A. Boccaletti, \email{anthony.boccaletti@obspm.fr} }


\authorrunning{A. Boccaletti et al.}
\titlerunning{NZ Lup}

\abstract
        {Planetary systems hold the imprint of the formation and of the evolution of planets especially at young ages, and in particular at the stage when the gas has dissipated leaving mostly secondary dust grains.  
        The dynamical perturbation of planets in the dust distribution can be revealed with high-contrast imaging in a variety of structures.  }
        {SPHERE, the high-contrast imaging device installed at the VLT, was designed to search for young giant planets in long period, but is also able to resolve fine details of planetary systems at the scale of astronomical units in the scattered-light regime. As a young and nearby star, NZ Lup was observed in the course of the SPHERE survey. A debris disk had been formerly identified with HST/NICMOS.}
        {We observed this system in the near-infrared with the camera in narrow and broad band filters and with the integral field spectrograph. High contrasts are achieved by the mean of pupil tracking combined with angular differential imaging algorithms.}
        {The high angular resolution provided by SPHERE allows us to reveal a new feature in the disk which is interpreted as a superimposition of two belts of planetesimals located at stellocentric distances of $\sim$85 and $\sim$115\,au, and with a mutual inclination of about 5$\degb$. Despite the very high inclination of the disk with respect to the line of sight, we conclude that the presence of a gap, that is, a void in the dust distribution between the belts, is likely.}
        {We discuss the implication of the existence of two belts and their relative inclination with respect to the presence of planets.} 
\maketitle

\section{Introduction}

Debris disks correspond to a late stage in the evolution of planetary systems when the primordial material has been expelled out of the systems or incorporated into planets and other bodies. The observed dust results from collisions among the rocky planetesimals or is deposited by comets. 
Planets, if already formed, are expected to produce indirect signatures in the form of a departure from a pure axisymmetrical disk morphology. 
Recently, the advance of high-contrast imaging, in particular with the installation of SPHERE \citep[Spectro-Polarimetic High contrast imager for Exoplanets REsearch,][]{Beuzit2019} and GPI \citep[Gemini Planet Imager,][]{Macintosh2008}, has yielded a significant number of new discoveries in this field either revealing new disks \citep{Lagrange2016, Kalas2015, Currie2015, Bonnefoy2017, Sissa2018} or new structures in known disks \citep{Boccaletti2015, Perrin2015, Garufi2016, Perrot2016, Milli2017}. These observations are definitely pointing to the presence of planets. 

Of particular interest is the ever-growing number of disks in which multiple belts are observed due to significant gain in angular resolution and contrast, both in the thermal emission and scattered light regimes.  
On the one hand, the sub-millimeter interferometer ALMA (Atacama Large Millimeter Array) has revealed obvious cases of gaps in several protoplanetary disks likely sculpted by sub-Jupiter-like planets \citep{Dipierro2015, Nomura2016}, as well as in one debris disk \citep{Marino2018}. 
On the other hand, at shorter wavelengths in scattered light, gaps were also found in the  protoplanetary disks of TW\,Hya \citep{Rapson2015b, vanBoekel2017} and V4046\,Sgr \citep{Rapson2015}, for instance. 
A few cases featuring an alternance of gaps and belts were also observed in some debris disks such as HD\,131835  \citep{Feldt2017} and HD\,141569 \citep{Perrot2016}. However, these disks contain gas \citep{Kral2017}, which might also be responsible for developing  belts \citep{Takeuchi2001,Kral2018} or arcs \citep{Lyra2013, Richert2018}. Despite the fact that a double-belt structure has been inferred for several systems from analyses of spectral energy distribution (SED) \citep{Pawellek2014, Pawellek2015}, so far a single gasless debris disk featuring two belts has been unambiguously imaged around HIP\,67497 by \citet{Bonnefoy2017}.
This potentially new class of debris disks is different from the systems with inner (warm) and outer (cold) components as inferred from photometric measurements at mid- and far-IR \citep{Chen2014}. Instead the multiple belts we see in scattered light with high-contrast imaging are all located at several tens of astronomical units, hence rather cold. It is tempting to hypothesize that multiple-belt systems could be one particular stage in the history of debris disks, later evolving as single-belt systems with a broad inner depleted cavity once all the planets have cleared out their orbits. Therefore, the moment when these substructures form and when they disappear is crucial to understand the formation and architecture of planetary systems. Observing one specific system provides a single snapshot in a disk lifetime. To capture the ``big picture'', several systems with various morphologies and at different stages of evolution must be found and studied. 

The star NZ Lup (HD\,141943, TYC\,7846-1538-1 , G2, V=7.97, H=6.41) is known to harbor a debris disk first inferred from {\it Spitzer} photometry \citep{Hillenbrand2008}. \citet{Gaia2018} measured a distance of $60.34^{+0.19}_{-0.18}$\,pc.
The  IR excess in the SED is modeled as two blackbodies peaking at equivalent temperatures of 197\,K and 60\,K and  corresponding to physical separations of respectively 4 and 122\,au \citep[assuming d=67\,pc,][]{Chen2014}. The cold component is in fact poorly constrained with just a single photometric point measured by Spitzer/MIPS at $70\muup$m for which $F_{70\mu m}/F_*=15.66\pm2.3$ \citep{Hillenbrand2008}.
As a young star (see Section \ref{sec:star}) it has been a target for exoplanet searches by direct imaging \citep{Chauvin2010,Galicher2016,Vigan2017} but none of these studies reported hints of a disk in scattered light. 
While the inner belt is presumably too close to the star ($\sim$67mas) to be resolved, the outer belt was finally detected in NICMOS/HST (Near Infrared Camera and Multi-Object Spectrometer/Hubble Space Telescope) archival data in which it appears nearly edge-on ($i=85\degb$) with a maximum intensity of $\sim$0.25\,mJy.arcsec$^{-2}$ \citep{Soummer2014}. The angular resolution of NICMOS did not permit confirmation of the physical size of the belt inferred from the SED, but the signal of the scattered light is detected from 0.7$''$ to 2.5$''$.
Using the high-precision spectrometer HARPS, \citet{Lagrange2013} found no planet more massive than 1-5\,M$_J$ for periods shorter than $\sim$100 days ($\sim$0.4\,au). 

In this paper we present the discovery of two cold belts in the debris disk of NZ Lup. The characteristics of the star are presented in Section \ref{sec:star}, while the observations and data reductions are provided in Section \ref{sec:obs}. A general description of the disk morphology is presented in Section \ref{sec:description}, and this geometry is studied in more detail using modeling in Section \ref{sec:model}.   The SED is revisited in Section \ref{sec:sed}. We provide the astrometric characterization of the point sources contained in the field of view as well as the estimation of the  limits of detection in Section \ref{sec:contrast}. Finally, we discuss the implications of  double belt structure with respect to the presence of planets (Section \ref{sec:discussion}). 


\section{Host-star properties}
\label{sec:star}

The object NZ\,Lup is a bona fide young star (age $<$\,50 Myr), as revealed consistently by a variety of indicators (e.g., strong lithium line, fast rotation, strong magnetic activity).
It lies in front of the Upper Centaurus Lupus (UCL) group and shares similar kinematic parameters. For this reason, it was proposed as a UCL member (age 17 Myr) by \cite{Song2012}.

There are indications of a modest amount of reddening (E(B-V)=0.03-0.05), comparing the expected colors for a G2 star \citep[spectral type from][]{Torres2006}  to the \cite{Pecaut2013} sequence for young stars.

\begin{center}
\begin{figure}
\includegraphics[width=9cm, trim= 2cm 4cm 2cm 11cm , clip]{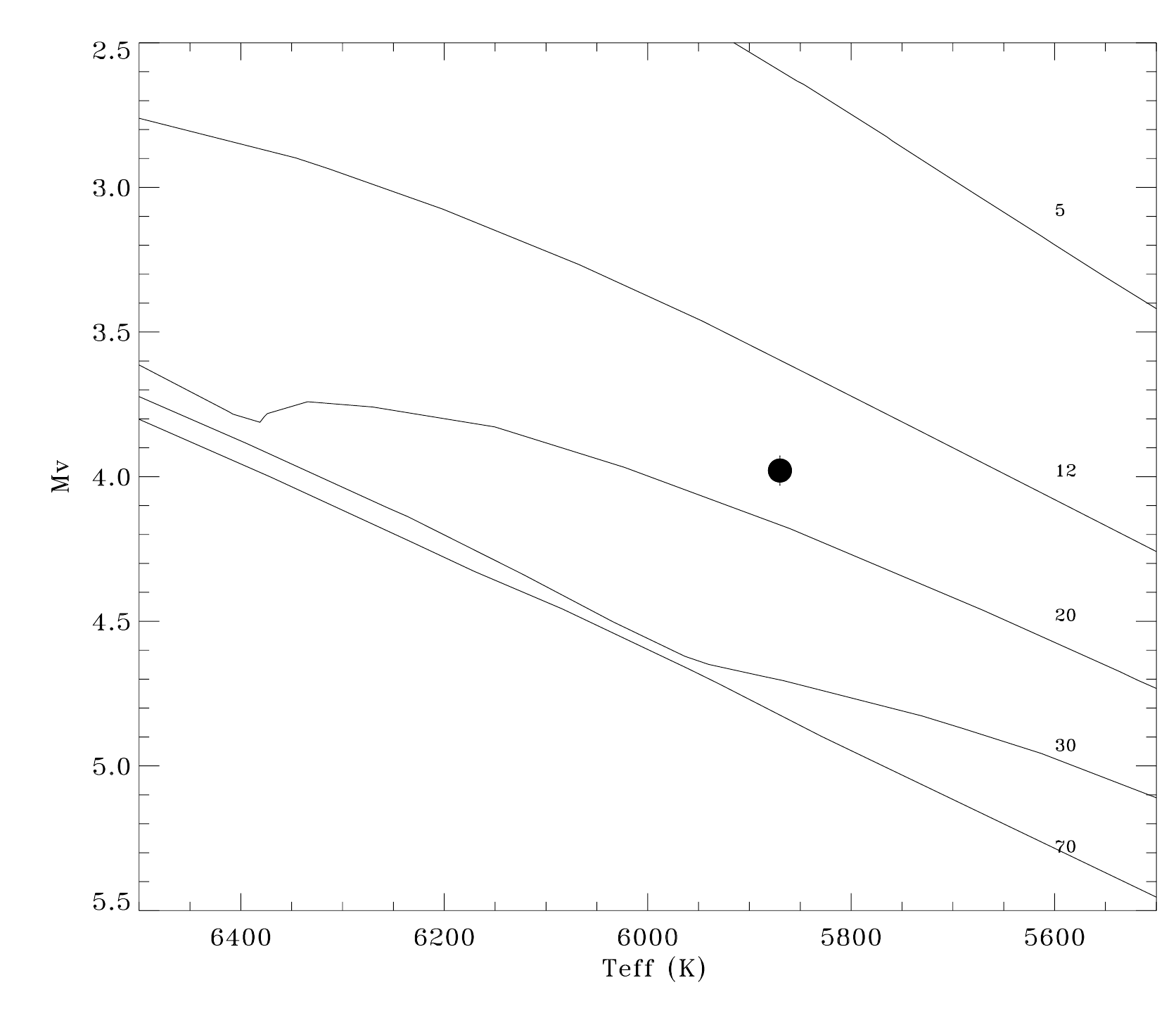}
\caption{Color-magnitude diagram of NZ Lup. Overplotted isochrones for 5-12-20-30 
and 70 Myr from \citet{Bressan2012}.} \label{f:isoc}
\label{fig:cmd}
\end{figure}
\end{center}

A robust isochrone age determination was prevented recently by a lack of trigonometric parallax. Exploiting Gaia DR2 \citep{Gaia2018} parallax and coupling it with effective temperature from spectral type (5870 K) 
and observed magnitudes in different bands shows that the star has not yet settled on the main sequence and is therefore very young (Table \ref{tab:star} and Fig. \ref{fig:cmd}).  
Comparison of the stellar parameters from Table \ref{t:param} (including the small amount of reddening we derived) with the models of \citet{Bressan2012} yields an age of 16$\pm$2 Myr. \citet{Marsden2011} mentioned the possibility of binarity, because of a marginal radial-velocity variability. Unresolved binarity could explain the off-sequence position on the color-magnitude diagram (CMD).
However, higher-precision measurements by \citet{Lagrange2013} with HARPS (High Accuracy Radial velocity Planet Searcher) indicate only a moderately large scatter linked to magnetic activity, as demonstrated by the RV-line bisector correlation.
An analysis of SPHERE images presented in the following section (including the noncoronagraphic ones taken during target acquisition) is also missing any evidence of multiplicity down to a very small separation of about 40 mas (2.4\,au). We therefore dismiss the possibility of binarity and consider the derived isochrone age
to be reliable.

The isochrone age is further supported by indirect indicators, such as lithium and magnetic/coronal activity, that are broadly compatible with ages of between $10$ and $50$ Myr. The rotation period is slightly longer than that of stars of similar color in the $\beta$ Pic moving group \citep{Messina2017}, consistent with the slightly younger age derived with isochrone fitting.

Finally, the kinematic parameters are very similar to the UCL ones, although the star lies at a shorter distance than the bulk of UCL members  (about 140\,pc). The Banyan $\Sigma$ online tool \citep{Banyan2018} yields a  probability of 55\% of UCL membership, and no significant membership probability for other groups. We therefore conclude that NZ Lup is a star with UCL age and kinematics but that it is in front of the main body of the group. A link between this target and the UCL seems probable but  the evaluation of the actual extension of the Sco-Cen groups at distances much smaller than 100\,pc is beyond the scope of the present paper. We therefore adopt an age of 16 Myr. 
\begin{table}[t]
\begin{center}
\caption{Stellar parameters of NZ\,Lup}\label{t:param}
\begin{tabular}{lcl}
\hline\hline
Parameter      & Value  & Ref \\
\hline
V (mag)                   &    7.975    & \cite{Kiraga2012}  \\
B$-$V (mag)               &    0.65   &  \cite{Torres2006}  \\
V$-$I (mag)               &    0.762         & \cite{Kiraga2012}  \\
J (mag)                   &     6.738$\pm$0.024  & 2MASS \\
H (mag)                   &     6.413$\pm$0.026  & 2MASS \\
K (mag)                   &     6.342$\pm$0.026  & 2MASS \\
Parallax (mas)            &    16.5716$\pm$0.0507 & \citet{Gaia2018} \\  $\mu_{\alpha}$ (mas\,yr$^{-1}$)  & -43.084 $\pm$0.095  & \citet{Gaia2018} \\
$\mu_{\delta}$ (mas\,yr$^{-1}$)  & -65.518$\pm$0.068  & \citet{Gaia2018} \\
RV   (km\,s$^{-1}$)            &  -1.7$\pm$1.0  &  \cite{Torres2006} \\
ST                   &  G2    & \cite{Torres2006} \\
$T_{\rm eff}$ (K)      &  5870$\pm100$ & this paper \\
E(B-V)               &    0.04$\pm$0.01        & this paper \\
$v \sin i $  (km\,s$^{-1}$)         &   35.0$\pm$0.5 & \citet{Marsden2011} \\
$P_{rot}$             &     2.182         & \citet{Marsden2011} \\
$i_{*}$               &   70$\pm$10     & \citet{Marsden2011}  \\
$\log R_{HK}$         &    -3.95   & \citet{Isaacson2010} \\
$\log L_{X}/L_{bol}$   &    -3.40   &  this paper \\
EW Li (m\AA)         &    230.0   & \citet{Torres2006} \\
Age (Myr)            &    16$\pm$2 & this paper  \\
$M_{star} (M_{\odot})$   &    1.244$\pm$0.031  & this paper \\ 
$R_{star} (R_{\odot})$   &    1.411$\pm$0.043  & this paper \\
\hline\hline
\label{tab:star}
\end{tabular}
\end{center}
\end{table}

\begin{table*}[th!]
\begin{center}
\begin{tabular}{l l l r l r l r l r l r l r l r l r }
\hline
Date UT         &       prog. ID        &       Filter                  &         Fov rotation    &       DIT     &       N$_ \mathrm{exp}$       &       T$_\mathrm{exp}$        &     seeing                      &       $\tau_0$        &       TN                      \\
                        &                       &                               &       (\deg)          &       (s)     &                                       &       (s)                             &       ($''$)                          &       (ms)    &       (\deg)                 \\ \hline \hline
2016-05-25      & 097.C-0523    &       IRDIS - BB\_H   &       46.6                    &       64      &       64                              &       4096                            &       $0.92\pm0.10$           &       2.4     &       $-1.65$\\
2016-05-25      & 097.C-0523    &       IFS - YJ                &       46.8                    &       64      &       64                              &       4096                            &       $-$                             &       $-$     &         $-$\\ \hline
2016-05-31      & 097.C-0865    &       IRDIS - H2H3    &       46.0                    &       16      &       264                             &       4224                            &       $0.62\pm0.09$           &       3.6     &       $-1.72$\\
2016-05-31      & 097.C-0865    &       IFS - YJ                &       49.5                    &       64      &       72                              &       4608                            &       $-$                             &       $-$ &       $-$\\\hline
2017-04-30      & 198.C-0209    &       IRDIS - H2H3    &       60.3                    &       64      &       80                              &       5120                            &       $0.52\pm0.07$           &       5.0     &       $-1.78$\\
2017-04-30      & 198.C-0209    &       IFS - YJ                &       60.8                    &       64      &       80                              &       5120                            &       $-$                             &       $-$     &         $-$\\ \hline
\hline
\end{tabular}
\end{center}
\caption{Log of SPHERE observations indicating (left to right columns): the date of observations in UT, the ID of the ESO program, the filters combination, the amount of field rotation in degrees, the individual exposure time (DIT) in seconds, the total number of exposures, the total exposure time in seconds, the DIMM seeing measured in arcseconds, the correlation time $\tau_0$ in milliseconds, and the true north (TN) offset in degrees.} 
\label{tab:log}
\end{table*}

\section{Observations and data reduction}
\label{sec:obs}

SPHERE \citep{Beuzit2019} is the extreme AO \citep{Fusco2014} instrument installed at the VLT (Very Large Telescope, ESO-Chile) equipped with coronagraphs \citep{Boccaletti2008a}, which routinely delivers high-contrast imaging data of a large survey of young stars that began in 2015 \citep[SHINE, SpHere INfrared survey for Exoplanets,][]{Chauvin2017}.
We observed NZ Lup during guaranteed time on May 25, 2016, May 31, 2016 and Apr. 30, 2017, using the IRDIFS mode of SPHERE, in which both IRDIS, the Infra-Red Dual-band Imager and Spectrograph \citep{Dohlen2008}, and IFS, the Integral Field Spectrograph \citep{Claudi2008}, are operated simultaneously. The observing log is displayed in Table \ref{tab:log}. The first epoch was set up with a broad band filter (BB\_H) for disk-detection purposes, while the other two epochs were using narrow bands (H2=1.593$\muup$m, H3=1.667$\muup$m, $R\sim30$) for exoplanet-detection purposes \citep{Vigan2010}.   The IFS was configured in YJ mode (0.95-1.35$\muup$m, $R\sim54$).

The sequence of observations is as follows: 1) the target acquisition optimizes the position of the  star onto the coronagraphic mask to set up reference slopes for the wavefront sensor; 2) then the star is offset by $\sim0.5''$ from the mask to obtain a flux calibration (PSF); 3) the star image is sent back onto the mask and a waffle pattern is applied on the deformable mirror to create four satellite spots at a separation of 14\,$\lambda/D$ for centering the target in detector coordinates; 4) the waffle pattern is removed and the science exposures start while the detector is dithered on a 4$\times$4-pixel grid to further reject bad pixels; 5) both centering frames and flux calibration are repeated; 6) finally the AO loop is opened and the telescope moves to the sky background. 
All coronagraphic images were acquired with an APodized Lyot Coronagraph \citep[APLC, ][]{Soummer2005}, the focal mask of which is 185\,mas in diameter combined to an apodizer which transmits 67\% of the light \citep{Carbillet2011}.

The IRDIS and IFS data are reduced at the SPHERE Data Center\footnote{\url{http://sphere.osug.fr}} \citep{Delorme2017} using the SPHERE pipeline \citep{Pavlov2008} and following a standard cosmetic reduction (sky subtraction, flat field correction, bad pixel removal). Raw frames are corrected for distortion \citep{Maire2016}. The star position in the image is determined from the satellite spots as detailed in \citet{Boccaletti2018} and no further recentering is performed. More details are provided in \citet{Boccaletti2018}. The north orientation is calibrated with astrometric reference fields \citep{Maire2016}. The IRDIS and IFS pixel scales are 12.25\,mas and 7.46\,mas, respectively. 

Starting from the output of this reduction, the four-dimensional data cubes (spatial, spectral and temporal dimensions) were processed with \texttt{SpeCal}, the differential imaging implementation at the SPHERE data center \citep{Galicher2018}. Several types of angular differential imaging (ADI) techniques  were considered \citep{Marois2006,Lafreniere2007,Marois2014,Soummer2012}. The study presented here is based on the principal component analysis \citep[the KLIP algorithm,][]{Soummer2012}, but other algorithms provide similar images. In practice, the resulting KLIP image depends on the number of the lowest-order modes that are kept to build a reference frame, which acts as a spatial filtering of low-frequencies to reject stellar residuals. 

\begin{figure*}[ht!] 
\centering
\includegraphics[width=18cm]{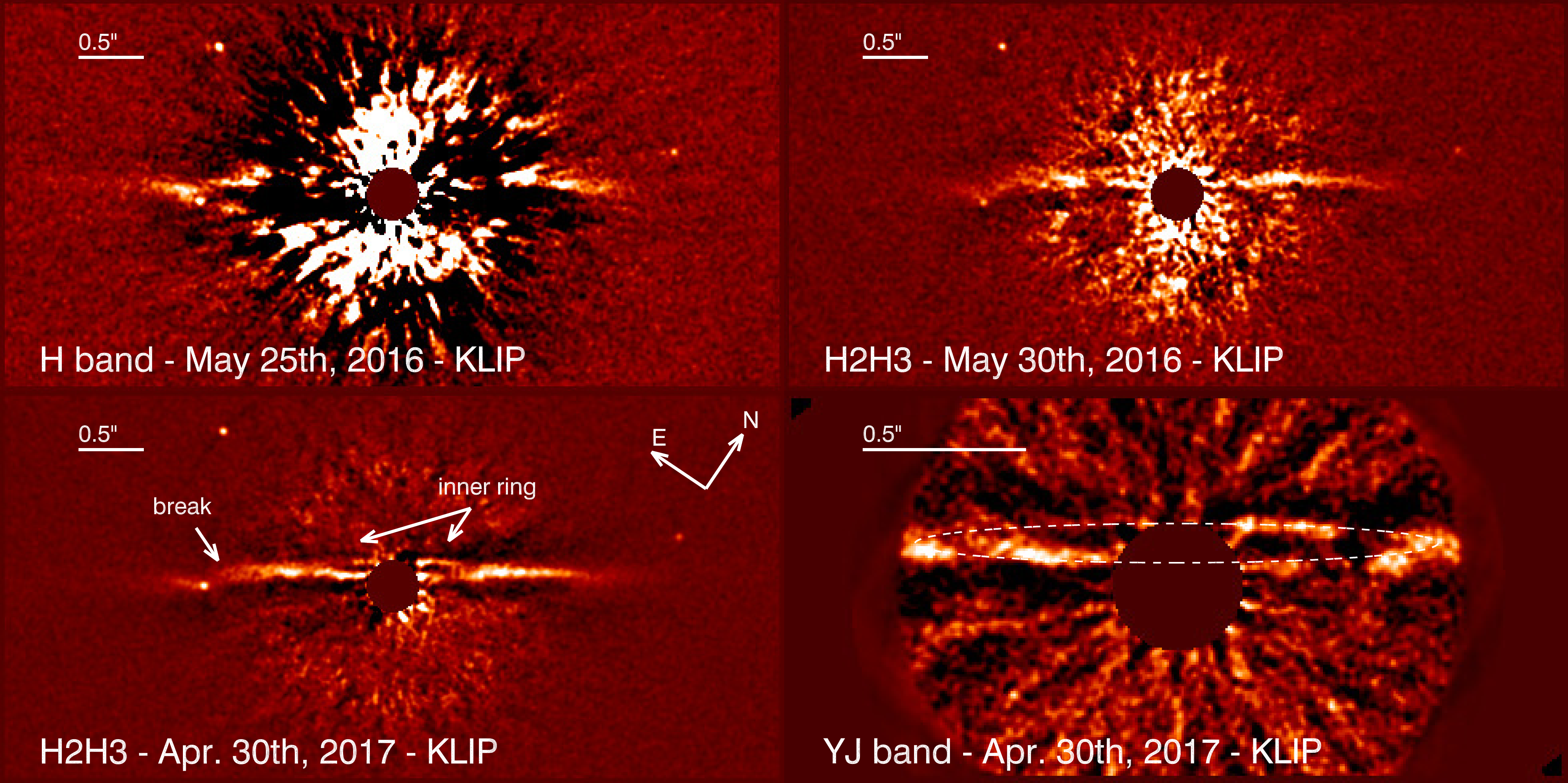}
\caption{Images of the disk obtained with IRDIS in the H band (top and bottom-left), and with the IFS in the YJ band in a narrower field of view (bottom-right). The intensity scale is arbitrary for each epoch to optimize the disk visibility. The disk midplane is aligned with the horizontal direction and the displayed field of view is $6''\times3''$ for IRDIS and $2.4''\times1.2''$ for the IFS. The main features reported in the text are annotated with arrows and ellipse.}
\label{fig:images}
\end{figure*}

\section{General description}
\label{sec:description}

The images processed with ADI using the KLIP algorithm are displayed in Fig. \ref{fig:images}. The top panel presents the two IRDIS observations from May 2016, while those from April 2017 are shown for both IRDIS and IFS at the bottom.  The best-quality data are achieved for the last epoch, as is obvious from the images; these correspond to the best  seeing and coherence time parameters (Table \ref{tab:log}) as well as the largest contrasts (see Section \ref{sec:contrast}). 
The reference frames that are subtracted out from the data set in the KLIP procedure were obtained with 10 modes for all reductions, except the one from May 25, 2016 (IRDIS), which used 40 modes. This larger number of modes (compared to the amount of frames; see Table \ref{tab:log}) is the direct consequence of lower-quality data and a broader stellar halo. The disk analysis below is based on the April 2017 data. 

The disk is oriented southeast to northwest at a position angle ($PA$) of $146.53\pm0.15\degb$, measured following the procedure described in \citet{Boccaletti2018}, in which the error bar includes the measurement uncertainty ($\sim0.1\degb$) together with the TN uncertainty ($\sim0.1\degb$).
The general aspect of the disk corresponds to a very inclined ring ($\sim 85 \degb$) of which the ansae are located at about  $1.3''$ southeast and $1.4''$ northwest, while only one single side is visible. This bright side should presumably be the front side if forward scattering is predominant as expected for small dust grains. 
Although the main disk stops at $\sim1.5''$ on both sides, we see a signal of dust scattering in the disk direction out to $\sim2.2''$ on both sides. 
This global morphology agrees well with the NICMOS image although a higher angular resolution is achieved with SPHERE \citep{Soummer2014}.

A closer examination of the disk reveals two unusual characteristics. First the southeastern ansae features a {\it break}, coincident with the location of a point source (indicated with an arrow in Fig. \ref{fig:images}). This object was removed in each frame of the data cube by subtraction of a scaled PSF and the break is still observed, indicating that it is not induced by an ADI artifact caused by the overlap of a point source on top of the disk image. In any case, this point source is not related to the system but is flagged as a background star (Section \ref{sec:contrast}), and so cannot be responsible for a dynamical effect on the disk. 

\begin{figure}[t!] 
\centering
\includegraphics[width=9cm]{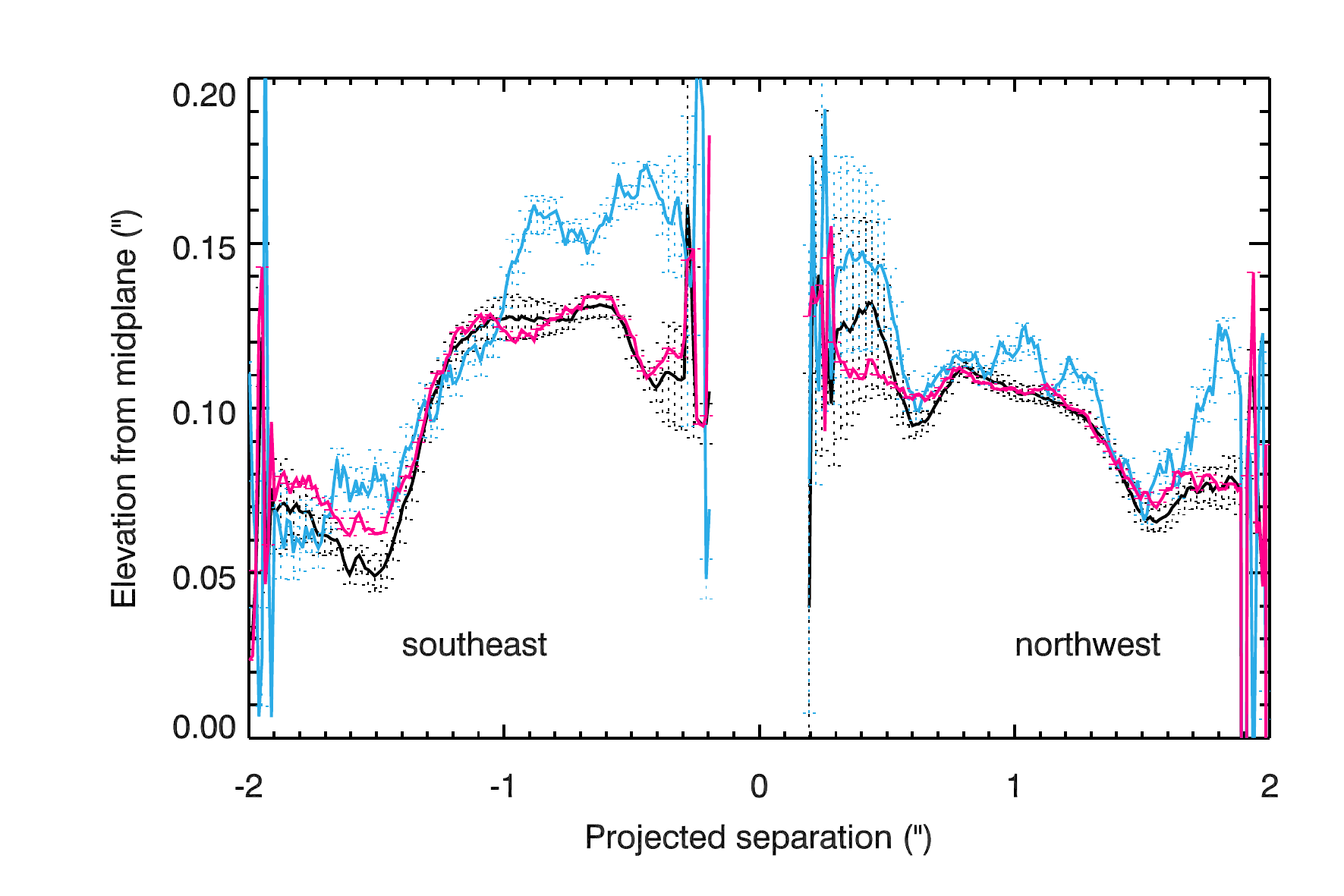}
\caption{Spine of the disk measured with a single component assumption (black line) and for two components (red and blue lines).}
\label{fig:spine}
\end{figure}

Secondly, and most importantly, the disk splits in two parts at radii closer than $\sim0.6''$ as if it were an inner ring.
This disk splitting is detected unambiguously in both IRDIS and IFS images from April 2017 (arrows/ellipse in Fig. \ref{fig:images}, and Fig. \ref{fig:snr}, bottom) but is also identified in lower-quality data from May 2016 (Fig. \ref{fig:images} and Fig. \ref{fig:snr}, top right). We measured the spine of the disk (Fig. \ref{fig:spine}) by fitting a Lorentzian profile on the vertical cross section for each stellocentric distance, and considering either a single component or two components (to account for the disk splitting mentioned above). The single-component fit (black line) has two minima located at about 1.5$''$ on both sides, which correspond to the edges of the main disk, the one at the southeast being steeper than in the northwest. The maximum elevation with respect to the midplane is about $0.13''$. The fit with two components (blue and red lines) was ordered according to the intensity of the component (the brightest being the closest to the midplane). The brightest component (red line) agrees well with the single component especially in the southeast. The faintest (blue line) deviates from the main spine starting at 1$''$ from the star and inwards while it culminates at $\sim0.17''$. This departure comes in fact with decreased elevation of the brightest component inwards of $\sim0.6''$ (in accordance with the ring-like shape in the image).
The spine is noisier in the northwest, making it more difficult to distinguish the two components in Fig. \ref{fig:spine}. Overall, the measurement of the spine quantifies the main pattern that is seen in the image, where the inner ring appears superimposed on the main disk.

While the images convey the idea that a second ring would be sitting on top (higher elevation from the star) of the main ring, one should consider that the actual distribution of dust is altered by the ADI process, and cross talks are to be expected for intricate geometry. In any case, such a configuration would be difficult, if not impossible, to explain dynamically. Instead, we posit that the disk of NZ Lup is composed of two belts of different sizes, separated by a gap, and mutually inclined by a few degrees. Such an assumption was already suggested for the very inclined debris disk of HD\,15115 \citep{Engler2018}. The following section is dedicated to the modeling of this structure in order to test our hypothesis.

\begin{table*}[t!]
\caption{Best parameters and dispersion for several one- and two-belt models. The dispersion is not provided when a parameter value is taken as a prior (boldface) or when the frequency plot has a single peak.  For the two-belt models, we provide the mean values and dispersions (when relevant) from the frequency plots, as well as the best model values (in bracket).}
\begin{center}
\begin{tabular}{lcccccccccc} \\ \hline
model                           & i$_1$ [$^\circ$]      & $r_{1}$ [au] & i$_2$ [$^\circ$]            & $r_{2}$  [au]             & $\alpha_{in-1}$             & $\alpha_{out-1}$ & $\alpha_{in-2}$          & $\alpha_{out-2}$    &  $g$    & $h$\\ \hline\hline
\multicolumn{11}{c}{\bf{1 belt}}\\ \hline
\texttt{mask1}          &  -                    & -                             &       85                             & $92\pm2$                        &    -                           &        -              & $9\pm2$        &        $-4.5\pm0.5$  & $0.6\pm0.05$ &  $0.015\pm0.005$ \\
\texttt{mask05}         &  -                    & -                             &       $85 \pm0.5$       & $93\pm2$                        &    -                          &        -               & $7\pm3$       &         $-5.0\pm0.5$  & 0.6             &      $0.020\pm0.005$\\
\texttt{mask03}         &  -                    & -                             &       $85 \pm0.5$       & $93\pm2$                        &    -                          &        -               & $5\pm6$       &        $-4.5\pm0.5$   & 0.6             &      $0.025\pm0.005$\\ \hline
\multicolumn{11}{c}{\bf{2 belts}}\\ \hline

\texttt {fill}             &       $82.7\pm0.6$    (83)    &       $89\pm4$ (90)    &       87      &       $106\pm5$ (105) &       4              &  \bf{-3.5}        &       -               & -4  & \bf{0.6}   &       $0.02$          \\      
\texttt {ring}          &       $82.4\pm2.0$    (82)    &       $89\pm4$ (90)    &       86      &       $ 90\pm10$ (90)  &     \bf{100} 	 &\bf{1}           	&       -               & -4  & \bf{0.6}    &       $0.02$          \\
\texttt {gap}          &       $81.5\pm0.6$    (82)    &       $88\pm2$ (90)    &       87      &       $115\pm6$ (115)  &      4               &\bf{-20} 		&\bf{20}  		& -4   & \bf{0.6}        &       $0.01$  \\
\texttt {2gaps}      &       $82.3\pm0.7$    (82)    &       $87\pm5$ (85)    &       87      &       $116\pm5$ (115)  &     \bf{20}        &\bf{-20} 		&\bf{20}  		& -4   & \bf{0.6}        &       $0.02$  \\

\hline
\end{tabular}
\end{center}
\label{tab:bestmodels}
\end{table*}%

\section{Modeling of the belts}
\label{sec:model}

Given that the structures observed in the NZ\,Lup disk are rather fine and relatively faint, we proceed in several steps for the modeling. We first consider a single belt scenario in Section \ref{sec:model1b}, where we present the global assumptions to produce scattered light images of synthetic disks. We analyze the residuals between this one-belt model and the actual image to motivate a more refined analysis including two belts. In Section \ref{sec:model2b}, considering that the two-belt model has a rather large number of parameters in regards of the S/N, we define density functions to explore the distribution of the dust in between the two belts more specifically. We firstly assume that the outer belt is more inclined than the inner belt with respect to the line of sight as it appears more relevant from the image. We then check that this hypothesis is valid for one single type of model in Section \ref{sec:model2breverse}.

\subsection{One-belt scenario}
\label{sec:model1b}

We used a simplified version of {GRaTer} \citep{Augereau1999} to produce synthetic images of debris disks with no particular assumption about the grain composition. 
We assume that dust is collisionally produced from parent bodies located in an axisymmetric narrow birth ring, 
at a separation $r_0$ from the star at which we assume a density $n_0$, the maximum density in the disk.  The edge of the ring observed at about 1.5$''$ corresponds to approximately 90\,au for a distance of 60\,pc. We further assume that the parent belt is small with respect to the angular resolution, and that the dust seen outside the parent belt corresponds to very small grains placed there either by PR-drag (in the $r<r_0$ region) or by high-eccentricity orbits induced by radiation pressure (in the $r>r_0$
region) \citep{Strubbe2006, Thebault2008}.

The density function is modeled by three components as follows. 
\begin{equation}
n(r,z) \propto n_0.R(r).Z(r,z) 
\end{equation}

The radial profile function $R(r)$, with $r$ being the radial dimension in the disk plane, is described by power laws with a maximum at the location of the belt and decreasing inward ($\alpha_{in}>0$) and outward ($\alpha_{out}<0$).

\begin{equation}
R(r) = \left( \left( \frac{r}{r_0} \right)^{2\alpha_{in}} + \left(\frac{r}{r_0}\right)^{2\alpha_{out}} \right)^{-1/2} 
.\end{equation}

The vertical profile function ($Z(r,z)$, with $z$ being the vertical dimension perpendicular to the midplane, is assumed to be Gaussian, and the disk scale height ($h$) varies linearly with the radius ($h=H_0/r_0$, $H_0$ being the height of the disk at the position $r_0$). 

\begin{equation}
Z(r,z) = exp\left(  \left(-\frac{|z|}{h \times r}  \right)^2\right)
.\end{equation}

Finally, the scattering phase function  ($\theta$ being the scattering angle) is approximated analytically with the Henyey-Greenstein function controlled with the anisotropic scattering factor $g$.

\begin{equation}
f(\theta) = \frac{1 - g^2}{4\pi(1+g^2+2g\times cos(\theta))^{3/2} }
.\end{equation}

Assuming the scattering of dust particles is preferentially directed forward, $g$ is positive and greater than zero but smaller than one (isotropic scattering). The model is inclined with respect to the line of sight ($i$) and rotated in the sky plane ($PA$). We used a forward modeling approach which consists in generating a grid of models, processing them the same way as the data, and calculating a   $\chi^2$ metrics.  As a first guess, we started with a single belt geometry and generated a grid of 9600 models with the following parameters (with some priors regarding the acceptable range). \\ 
\smallskip\\
- position angle: PA  [$^\circ$] = 146.5 \\ 
- inclination: i [$^\circ$]  = 84, 85, 86\\
- radius of the belt: $r_0$ [au] = 65, 70, 75, 80, 85, 90, 95, 100 \\
- anisotropic scattering factor: g = 0.4, 0.5, 0.6, 0.7 \\
- slope of the surface density inward: $\alpha_{in}$ = 2, 4, 6, 8, 10\\- slope of the surface density outward: $\alpha_{out}$ = -2, -3, -4, -5, -6\\
- scale height: $h$ = 0.01, 0.02, 0.03, 0.04 \\

To compare the models with the data and to determine the best model parameters, we proceed as in \citet{Engler2018}.
Each model is convolved with the PSF, and projected onto the same KLIP basis as used to reduce the data in order to reproduce a similar level of self-subtraction on synthetic images. This new grid of models is the input of the minimization procedure. 
We define a rectangular aperture globally aligned with the disk axis to encompass the disk image, the  length of which is 6$''$ in total and is 0.4$''$ in  width. Because half of the disk (the northeast) is visible, the aperture is offset by 0.1$''$ with respect to the star along the minor axis to avoid including too much noise in the aperture. 
The reason the aperture extends out to 3$''$ although the main ring is located at $\sim1.5''$ is to take into account the scattering signal observed out to at least $2.2''$ on both sides, which is crucial to constrain the $\alpha_{out}$ parameter accounting for the halo of small grains beyond $r_0$.
The central part of the image is also removed numerically to hide the strongest stellar residuals. 
We tested three different cases for which the central masked region is 1.0$''$, 0.5$''$ , and 0.3$''$ in radius (\texttt{mask1, mask05, mask03}). 
However, since we are approximating a two-belt disk with a single belt model, we are mostly interested in the disk parameters at large separations in this first step.  
A $2\times2$-pixel binning is applied to the data and the models. 
For each model in the considered aperture, we derived the intensity scale which minimizes the quadratic difference between the data ($\mathcal{O}$) and the model ($\mathcal{M}$). We then obtained the reduced $\chi^2$  with the standard relation: 

\begin{figure}[t] 
\centering
\includegraphics[width=9cm]{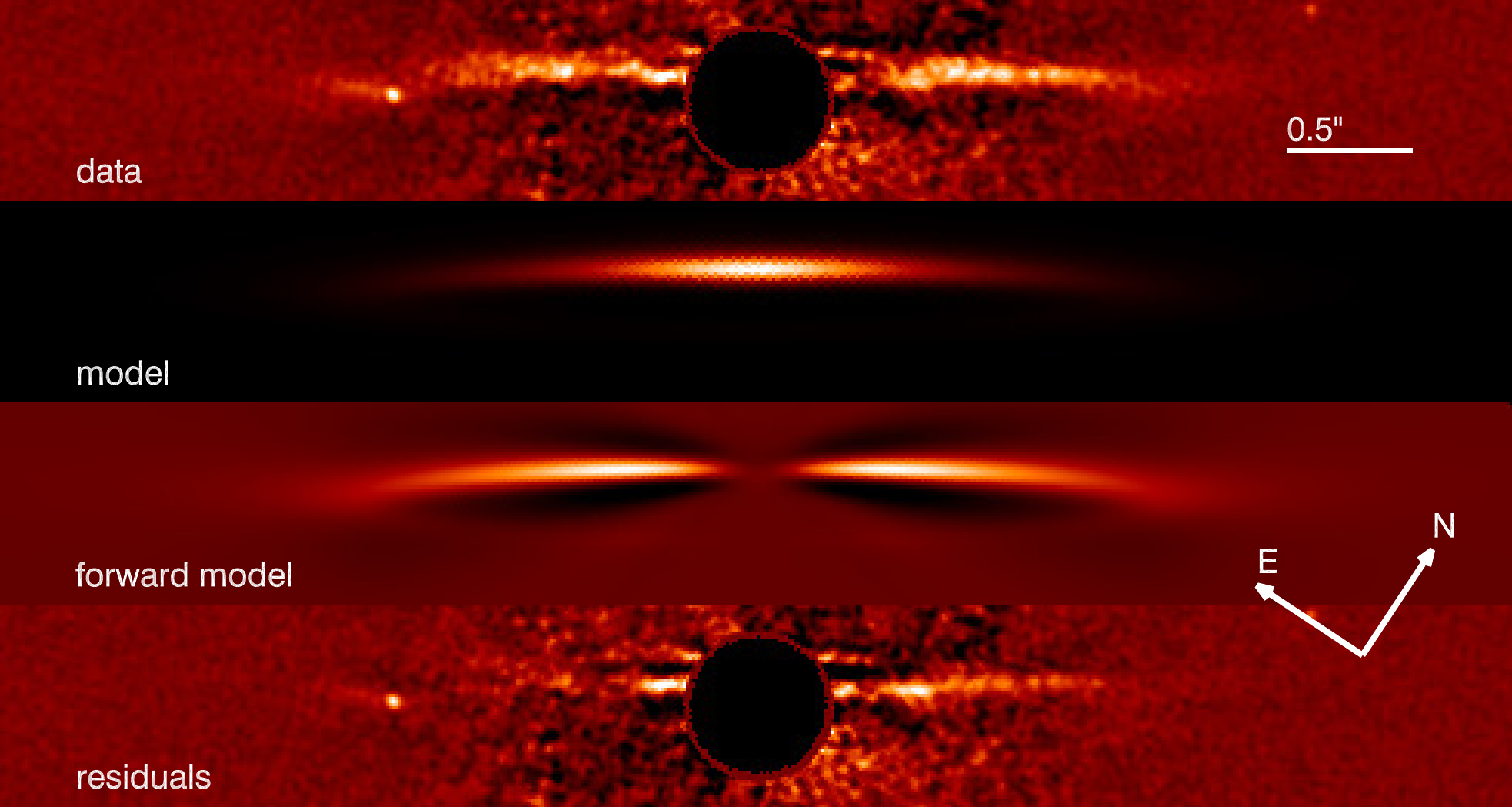}
\caption{From top to bottom: Original data from April 2017 (same as in Fig. \ref{fig:images}), the raw model before PSF convolution, the best forward model 
($i=85\degb$,  $r=95$\,au, $\alpha_{in}=10$, $\alpha_{out}=-5$, $g=0.6$, $h=0.02$) and the residuals. The field of view is $6''\times0.8''$.}
\label{fig:datavsmodels1b}
\end{figure}

\begin{equation}
\chi^2 = \frac{1}{\nu}\sum_{i,j}^{Ndata}{ \frac{\Big(\mathcal{O}(i,j) - \mathcal{M}(i,j)\Big)^2}{\sigma(i,j)^2}}
.\end{equation} 

The degree of freedom ($\nu$) is taken as $N_{data}-N_{params}$ with $N_{params}=6$ in this first case, with $N_{data}$ being the number of data points in the aperture after binning. The minimum value of $\chi^2$ provides the parameters for the best model. To determine the range of models which best matches the data in a conservative way, we considered a threshold at 1\% of the lowest  $\chi^2$ values instead of adopting the standard $\sqrt{2/\nu}$ threshold. The latter theoretically applies to Gaussian noise and linear models, two conditions not necessarily satisfied in the present case. The  parameter values and dispersions of the best-fitting models are provided in Table \ref{tab:bestmodels}, together with parameter frequencies of the 1\% best models in the Appendix (Fig. \ref{fig:freq_1belt}). The surface density slopes are consistent for all three cases suggesting a steeper density inwards than outwards ($\alpha_{in} > 5$, while $\alpha_{out}$ = -4 to -5). The latter is also steeper than the canonical value of -1.5 expected for a collisionally produced halo of small grains placed by radiation pressure beyond a main birth ring \citep{Strubbe2006, Thebault2008}.
However, the acceptable values of $\alpha_{in}$ can be very broad when the mask size decreases, so it is difficult make conclusions from this first approach. 
The best inclination is about  $\sim85-86^\circ$, the position of the planetesimal belt is at $\sim90-95$\,au, and the asymmetric scattering factor is $g\sim0.6$ indicating a rather high value of forward scattering 
\citep[but similar to other debris disks:][]{Lagrange2016, Olofsson2016, Bonnefoy2017}. Regarding the scale height, both $h=0.01$ and $h=0.02$ fit equally well at least in the case of model \texttt{mask1}.
Images of the model compared to the data, together with residuals, are shown in Fig. \ref{fig:datavsmodels1b}. The single belt is clearly unable to reproduce the disk splitting seen at stellocentric distances shorter than $1''$.

\begin{figure}[t!] 
\centering
\includegraphics[width=9cm]{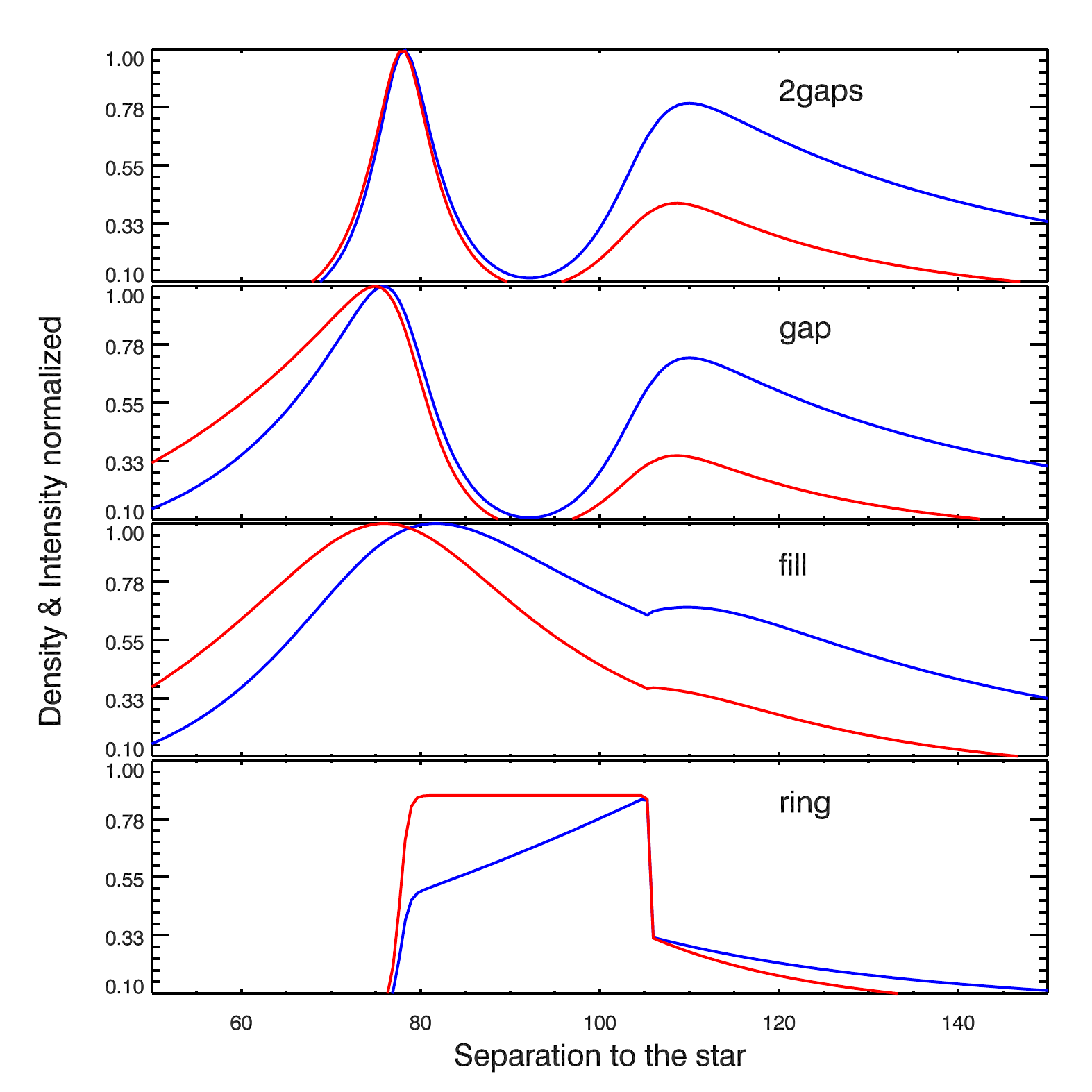}  
\caption{Examples of the density (blue) and intensity (red) functions corresponding to the four model families.}
\label{fig:sketch}
\end{figure}

\subsection{Two-belt scenario}
We now consider a more realistic configuration of two separate belts at stellocentric distances $r_1$ and $r_2$, and with independent parameters except for their relative inclinations. Due to the time-consuming nature of the forward modeling approach, we restrained the range of solutions by fixing some of the parameters. Following the single-belt modeling, we fix $g=0.6$ for the two belts. 
 The ADI process is known to significantly bias the images of disks, and in particular can emphasize the darkness of a central cavity for ring-like geometries. 
 Here, we are particularly interested in the density distribution in between the two belts, and therefore to investigate whether this region is filled or empty  we defined four families of radial densities by changing the surface density slopes ($\alpha_{in-1}$, $\alpha_{out-1}$, $\alpha_{in-2}$, $\alpha_{out-2}$\footnote{Subscripts 1 and 2 denote the first and second belt, respectively, with respect to the distance to the star.}): \texttt{fill}, \texttt{ring}, \texttt{gap}, \texttt{2gaps}. 
The values of the density slopes are provided in Table \ref{tab:bestmodels} and an illustrative sketch is depicted in Fig. \ref{fig:sketch}.
For the \texttt{gap} and \texttt{2gaps} cases the density slopes between $r_1$ and $r_2$ were chosen to be steep enough ($+/-$20) to avoid mutual overlapping. In addition, the \texttt{2gaps} models feature an additional inner gap ($r<r_1$).
For the \texttt{fill}  and \texttt{ring} cases the outer belt is truncated inwards of $r_2$. In the \texttt{fill} models the outer density slope of the first belt is fixed at -3.5 to avoid discontinuities between the belts. However, this slope is an average value since it does not match for any separations between $r_1$ and $r_2$.
Finally, the \texttt{ring} case features an increasing density function to allow a flat intensity variation. The outer disk density slope is the same for any model family ($-4$ or $-5$). In the case of \texttt{gap} and \texttt{fill} models, the inner slope is symmetrical with the outer one ($\alpha_{in-1}=-\alpha_{out-2}$).  Similarly, the scale height is allowed to take two values but be identical for the two belts ($h$ = 0.01, 0.02). 
 
\begin{figure}[t!] 
\centering
\includegraphics[width=9cm]{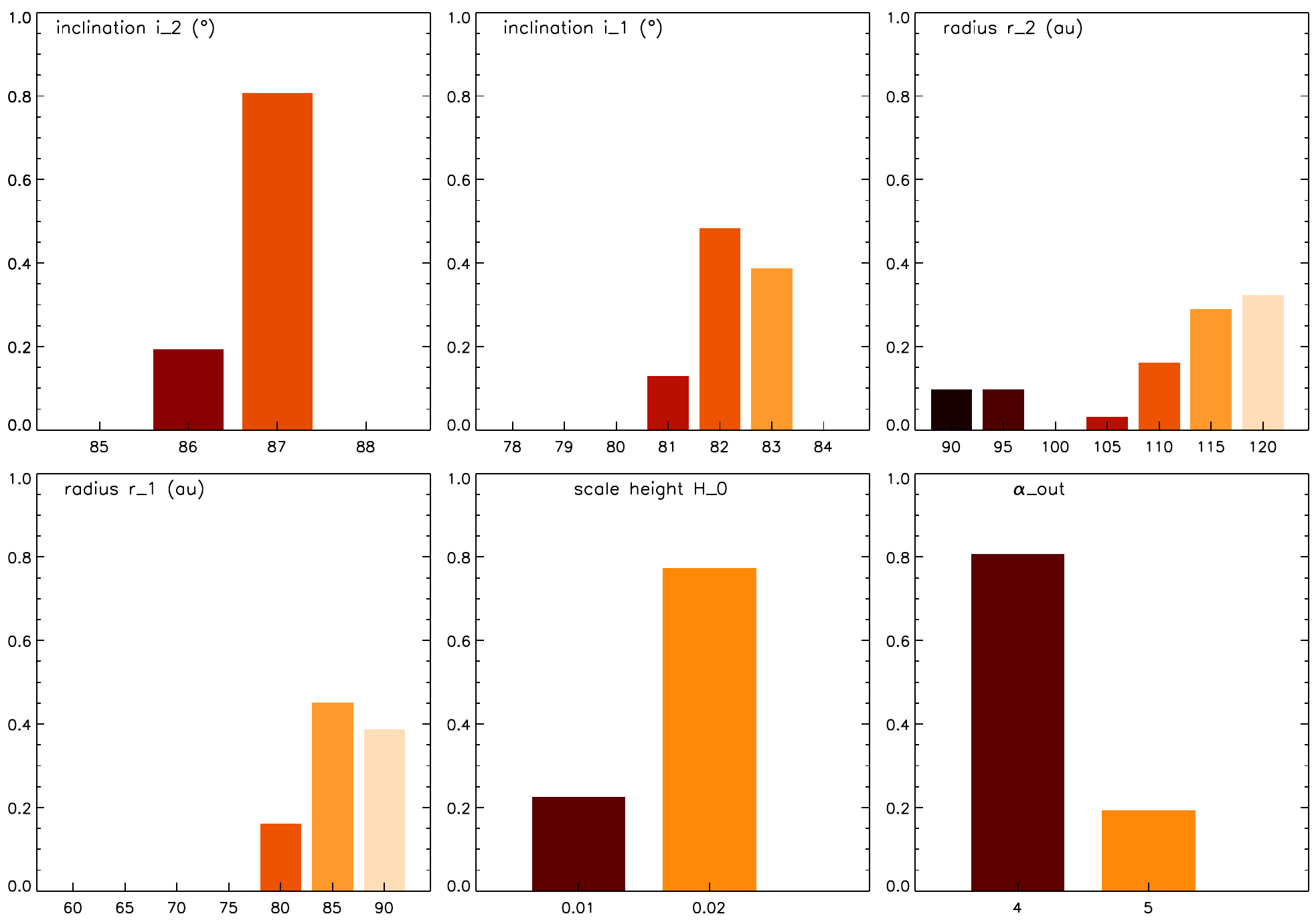}
\caption{Frequencies of parameters obtained for the 1\% best models in the two-belt \texttt{2gaps} case.}
\label{fig:freq_ultragap}
\end{figure}

 \subsubsection{Edge-on outer belt}
 \label{sec:model2b}

Once density functions are defined, the next parameters are the inclinations and the locations of the belts. Intuitively, from the visual inspection of the image we assume that of these two belts the most inclined should be the outer one. In that case, we defined the following ranges of parameters, again with some priors regarding the acceptable range:  
\smallskip\\
- inclination of the outer belt: $i_2$ [$\degb$]  = 85, 86, 87, 88 \\
- relative inclination $i_2-i_1$: [$\degb$]  =  4, 5, 6, 7,\\
- radius of the inner belt: $r_1$ [au] = 60, 65, 70, 75, 80, 85, 90  \\- radius of the outer belt: $r_2$ [au] = 90, 95, 100, 105, 110, 115, 120 \\

In the parameter space, we intentionally matched the largest value explored for $r_1$ with the minimal value explored for $r_2$ to check if unrealistic situations (same radii but different inclinations) could come out from the analysis. Here, we used GRaTer twice to model each belt individually and then the two synthetic images were co-added to produce a two-belt model. This is not exactly similar to defining a global analytical expression of the radial density but it allows more flexibility in adapting the relative weight and inclination of the belts (for optically thin disks the intensity is proportional to the density). 
Taking the phase function into account, we expect very strong forward scattering for very inclined disks, so a disk image in scattered light should be very strongly peaked at small phase angles. Furthermore, this peak quickly decreases in intensity as the scattering angles increase (or conversely the inclination decreases). In the two-belt configuration the less inclined (inner) belt therefore appears much fainter for a similar dust density to the more inclined (outer) belt, although it receives more stellar flux. This behavior is illustrated in Fig. \ref{fig:inclinationeffect}. However, we observe in the image a nearly identical intensity for these two belts, so implicitly  the inner belt should have a larger density to match the images.   
Qualitative tests led us to multiply the scattered light image of the inner belt by a factor two.%

\begin{figure}[t!] 
\centering
\includegraphics[width=9cm]{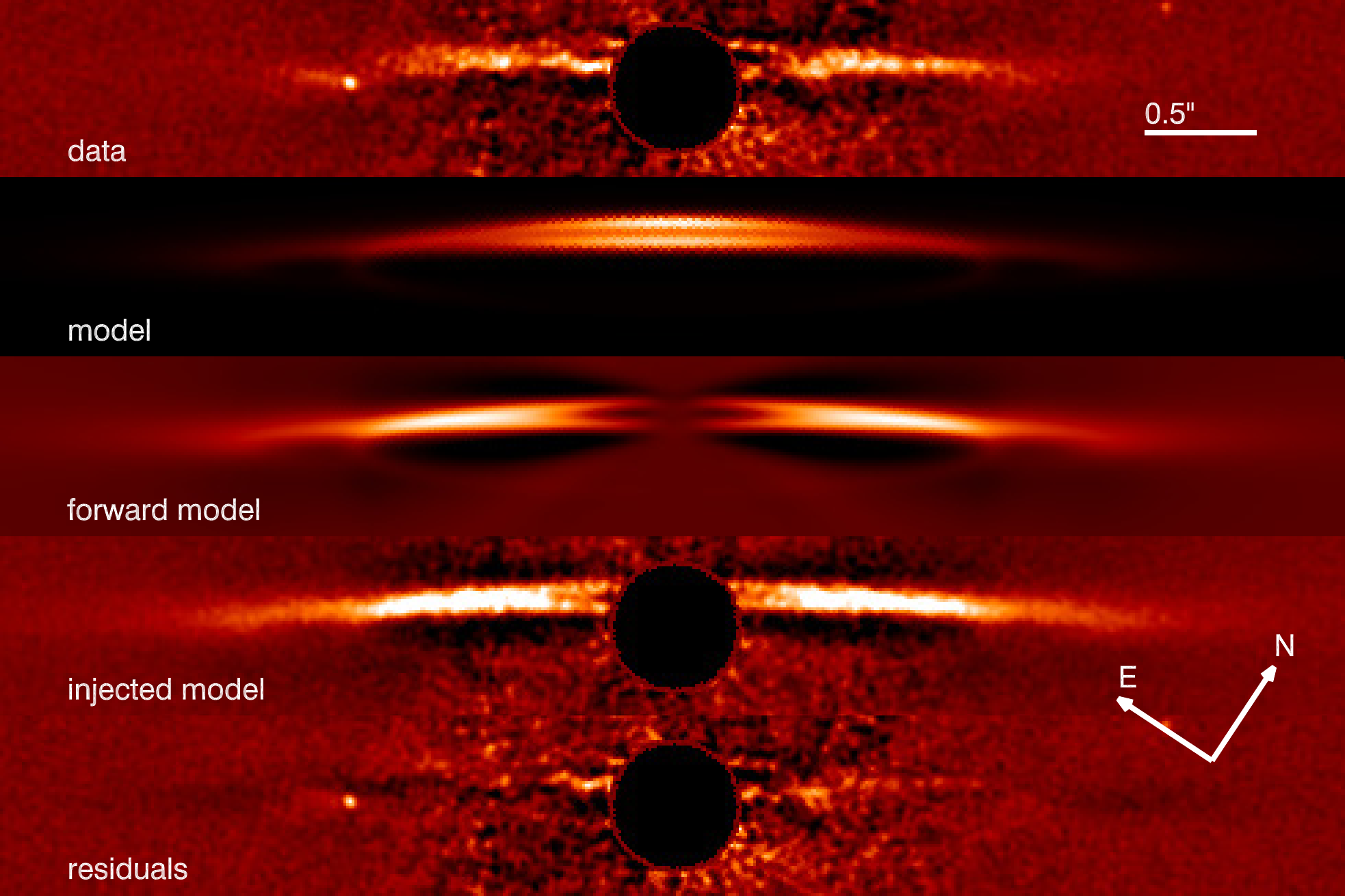}
\caption{From top to bottom: Original data from April 2017 (same as in Fig. \ref{fig:images}),  the raw model before PSF convolution, the corresponding forward model, the best model injected in the data, and the residuals (\texttt{2gaps} case: $i_1=82\degb$,  $i_2=87\degb$, $r_1=85$\,au,  $i_2=115$\,au). The field of view is $6''\times0.8''$.}
\label{fig:datavsmodels_ultragap}
\end{figure}

The aperture in which the $\chi^2$ is evaluated is similar to that of the single-belt case but reduced to 3$''$ instead of $6''$, the outer slope of the surface density being already constrained at the former stage. However, the choice of the aperture size has clearly some impact on the model fitting. A larger aperture tends to increase the $\chi^2$ and lead to similar solutions for all four of the families of models.
We tested two approaches: either a forward modeling, where, as before, a perfect model (no noise) is compared directly to the data, or alternatively,  the same model is injected into the data cube with a brightness ratio of $6\times10^{-6}$ (slightly brighter than the real disk), and processed the same way, but reversing the parallactic angle sequence to cancel out the disk while keeping the noise structure. 
In both cases, the eigenvectors of the PCA  are  determined on the data without any fake disk; these are then stored and reapplied on the noiseless model, or when the model is added to the data.  These two approaches yield mostly the same outcome and so the current results are based here on the latter for practical reasons. 

The $\chi^2$ for all tested models ranges from 4.15 to 11.75, globally indicating poor fits. 
The smallest $\chi^2$ are obtained for the \texttt{2gaps} family ($\chi^2_{min}=4.15$), then for  \texttt{ring} and \texttt{gap} ($\chi^2_{min}=4.32$ and 4.43 respectively), while  \texttt{fill} is clearly worse  ($\chi^2_{min}=4.64$). 
Overall, these values do not differ significantly, which means that the $\chi^2$  metric has some limitations to identify fine structures at low S/N in this particular case, while visually it can be straightforward to make a distinction between some models.
As in the one-belt scenario, we selected the best 1\% models for each of the four model families. The frequency of a parameter is defined as the occurrence of a value in the 1\% best models.  
Table \ref{tab:bestmodels} provides the most likely values for each parameter together with the dispersion estimated from a Gaussian fit of the frequency distribution when relevant (the parameter sampling is large enough and/or there is more than a single possible value). We do not provide a dispersion for $\alpha_{in-1}$, $\alpha_{out-2}$, and $h$ since only two values were tested. 
The images and the frequencies of each parameter for the best model family, \texttt{2gaps,} are displayed in Figs. \ref{fig:datavsmodels_ultragap} and \ref{fig:freq_ultragap}, respectively. Similar figures are shown in the Appendix for the other model families (Figs. \ref{fig:datavsmodels_truegap}, \ref{fig:datavsmodels_truefill}, \ref{fig:datavsmodels_justaring} and Figs.  \ref{fig:freq_truegap}, \ref{fig:freq_truefill}, \ref{fig:freq_justaring})
It is rather clear that only the best models for \texttt{2gaps} and  \texttt{gap} feature a disk splitting similar to the one observed in the data. However, there are some parameter combinations in the 1\% lowest $\chi^2$ for \texttt{fill} and  \texttt{ring} which can produce this pattern too.
 
For the inclination of the belts, the results are highly significant. The outer belt inclination ($i_2$) has a very high peak with frequencies as large as 80-100\% (\texttt{2gaps}, \texttt{gap}, \texttt{fill}), while for the inner belt ($i_1$) the significance is lower but still achieves frequencies larger than 50\%. The \texttt{ring} model family provides flatter histograms. Based on the minimum $\chi^2$ values, we obtained $i_1=82\degb$ and $i_2=87\degb$.
We observed more dispersion on the radii of the belts, which is a direct consequence of the disk being seen at  high inclination. If we exclude the \texttt{ring} family, which converges to $r_1=r_2=90\,au$, a likely unrealistic situation, then the external belt has its most likely position at about $r_2\sim105-115$\,au, while the inner belt is located at $r_1\sim85-90$\,au.  The frequency peak for $r_1$ and $r_2$  can be as small as $\sim$30\%, hence rather broad uncertainties of 5-10\,au. The best model family,  \texttt{2gaps}, provides the largest separation (30\,au $\approx 0.5''$) between $r_1$ and $r_2$, while the opposite is true for the  \texttt{fill} case (only 15\,au  $\approx 0.25''$).
As for the scale height, there is no strong prevalence of one value with respect to another except for the  \texttt{2gaps} model  which favors $h=0.02$.
This parameter is therefore not very well constrained in the two-belt analysis, but this is not very surprising since $h=0.02$ corresponds to a height of 2\,au (33\,mas) at a distance of 100\,au (about the position of the outer belt), which is also roughly the angular resolution of the images (40\,mas at $\lambda=1.6\muup$m). 
Contrary to the one-belt analysis, considering two belts in the system favors a shallower inner/outer slope of 4/-4 instead of 5/-5 although this is still compatible within error bars. 
We note that we observed degeneracies between the inclination of the inner belt ($i_1$) and its radius ($r_1$), in the sense that for the KLIP image to show a  disk splitting, this belt could be smaller ($\sim60-70$\,au) than what is obtained from the lowest $\chi^2$, but with a higher relative inclination with respect to the outer belt. However, such models are not part of any of the 1\% best models selected. We would need to increase the threshold to $\sim$5\% to start observing this degeneracy. 
Finally, the differences between the \texttt{2gaps} and \texttt{gap} cases are small enough to consider both of them as reliable descriptions of the disk image.

The intensity of the inner ring was multiplied by a factor of two as an initial guess and so this parameter was not included in the minimization. To provide further constraints on this intensity we now consider the best model
(\texttt{2gaps} case, $i_1=82\degb$, $i_2=87\degb$, $r_1=85$\,au  and $r_2=115$\,au, $h=0.02$, $\alpha_{out-2}$ = -4), and we vary the intensity weight of the inner belt from 1 to 3 (hence neglecting possible degeneracies between parameters). The optimal flux ratio is $1.8\pm0.1$ although it corresponds to a very small variation of $\chi^2$ compared to the former solution (4.13 instead of 4.15). In fact, the difference in the residual map is very difficult to appreciate by eye.

As a concluding note, there are strong limitations to the modeling of the images of the disk of NZ\,Lup   because of the high inclinations of the belts. The residual maps also show signs of asymmetries along the major axis which were not accounted for.  Still, we can reasonably conclude that the disk is made up of two belts with different sizes, inclinations, and densities. 

 \subsubsection{Edge-on inner belt}
\label{sec:model2breverse}

Another possible configuration not considered above corresponds to the opposite geometry, where the inner belt is at higher inclination (closer to edge-on) than the outer belt. We first considered the same grid as before, but changing $i_2$ with
$i_1$ and $i_2-i_1$ with $i_1-i_2$. The relative intensity is also adapted for the same reasons as provided in the previous section, so the outer belt synthetic image is multiplied by a factor of two. For simplicity, we only focused on the \texttt{gap} model and restrain the parameters  $\alpha_{in-1}$, $\alpha_{out-2}$ and $h$ to a single value (respectively 5, -5, and 0.02). The minimum $\chi^2$ for this grid of models is 9.81 (corresponding to $i_1=88\degb$, $i_2=83\degb$, $r_1=60$\,au, $r_2=90$\,au), much higher than in the previous configuration. In fact, the best models have a much thicker outer belt than in the data, and the apparent projected separation between the two belts is too large.  Therefore, we modified the range of inclination for the outer belt with $i_1-i_2=1,2,3,4\degb$. Even then however, the minimum $\chi^2$ is 6.22 (corresponding to $i_1=86\degb$, $i_2=85\degb$, $r_1=80$\,au, $r_2=95$\,au) showing again that these models do not provide a good match to the data.  In that case, the relative inclination is too small to generate an observable cavity between the  two belts as in the data. As a consequence, we can confidently rule out this geometry. 

 \subsubsection{Summary}

From the modeling work we were able to establish that the one-belt case does not  correctly match the data, leaving a significant residual in the inner part of the disk image. Therefore, the derived inclination and radius of the belt are biased and correspond to averaged values of a two-belt geometry. Using a two-belt scenario, we found that the presence of a gap in between the belts is a better match to the data. However, the differences between models with gap(s), a smooth transition, or no gap at all are small because the mutual inclinations and the distances of the belts could mimic a gap in the scattered light image. Some cases can still be rejected nevertheless, like for instance when the model converges to the same radii for the two belts when we consider a broad ring.  
Finally, models where the inner belt is more inclined than the outer belt can be confidently ruled out. 
In summary, the modeling favors a geometry with two belts at 85\,au and 115\,au, separated with a sharp gap, and  at inclinations of 85\deg and 87\deg, respectively.

\section{SED modeling}
\label{sec:sed}

\subsection{Stellar parameters and model setup}

Considering the identification of two belts in SPHERE observations, we decided to revisit the SED of the system to check for signs of such a structure. 
The contribution of the stellar photosphere to the total flux density is taken from a PHOENIX model \citep{Hauschildt1999} assuming a stellar luminosity of 2.9$\Lsun$ and a temperature of 6000~K \citep{Chen2014} for the host star. 
The photometric data were collected from several catalogs and from different papers summarized in Table~\ref{tab:photometry}.
The SED of the debris disk was fitted using the SONATA-code \citep{Mueller2009, Pawellek2014} using all data points longward of 10\,$\muup$m where we expect the dust excess emission to start, but the lack of data points in the far-IR (apart from those at 70 and 250\,$\muup$m) is clearly a limitation, preventing us from providing reliable constraints on the dust properties.

\begin{table}[t!]
\caption{Continuum flux density.
\label{tab:photometry}}
\tabcolsep 2pt
\begin{tabular}{llcrcc}
\hline
Wavelength      & \multicolumn{3}{c}{Flux density}      & Instrument & Reference                \\
$[\mum]$        & \multicolumn{3}{c}{[mJy]}             &            &                           \\
\hline \hline
1.235                   & 3174   &$\pm$& 70.16                                  & 2MASS           &       1                               \\
1.662                   & 2737   &$\pm$& 65.53                                  & 2MASS           &       1                               \\
2.159                   & 1902   &$\pm$& 45.55                                  & 2MASS           &       1                               \\
3.4                             & 876.6  &$\pm$& 68.63                                  & WISE            &       2                               \\
3.6                             & 849.1  &$\pm$& 18.2                                   & IRAC            &       3                               \\
4.5                             & 541.0  &$\pm$& 12.4                                   & IRAC            &       3                               \\
4.6                             & 528.2  &$\pm$& 13.14                                  & WISE            &       2                               \\
8.0                             & 193.1  &$\pm$& 4.2                                    & IRAC            &       3                               \\
9.0                             & 195.1  &$\pm$& 31.7                                   & AKARI   &       4                               \\
11.8                    & 93.41  &$\pm$& 1.46                                   & WISE            &       2                               \\
13                              & 100.36 &$\pm$& 2.54                                   & IRS             &       5                               \\
22.1                    & 32.54  &$\pm$& 1.53                                   & WISE            &       2                               \\
24                              & 28.11  &$\pm$& 0.57                                   & MIPS            &       5                               \\
31                              & 26.54  &$\pm$& 2.72                                   & IRS             &       5                               \\
33                              & 27.9   &$\pm$& 2.0                                    & IRS             &       3                               \\
70                              & 56.30  &$\pm$& 3.80                                   & MIPS            &       6                               \\
250                             & 32.7   &$\pm$& 13.6                                   & SPIRE           &       6                               \\
350*                    & 6.4    &$\pm$& 50.4                                   & SPIRE           &       6                               \\
500*                    & 19.0   &$\pm$& 45.6                                   & SPIRE   &       6                               \\
\hline
\end{tabular}

\noindent

{\em Notes:}
Asterisks give upper limits.
{\em References:}
[1] - 2MASS All-Sky Catalog of Point Sources; 
[2] - \cite{Wright2010}; 
[3] - NASA/IPAC Infrared Science Archive;
[4] - AKARI All-Sky Survey Bright Source Catalog;
[5] - \cite{Chen2014};
[6] - Herschel Science Archive
\end{table}
\begin{figure}[t!]
\centering
        \includegraphics[width=6.cm, trim= 2cm 2cm 2cm 2cm , clip, angle=-90]{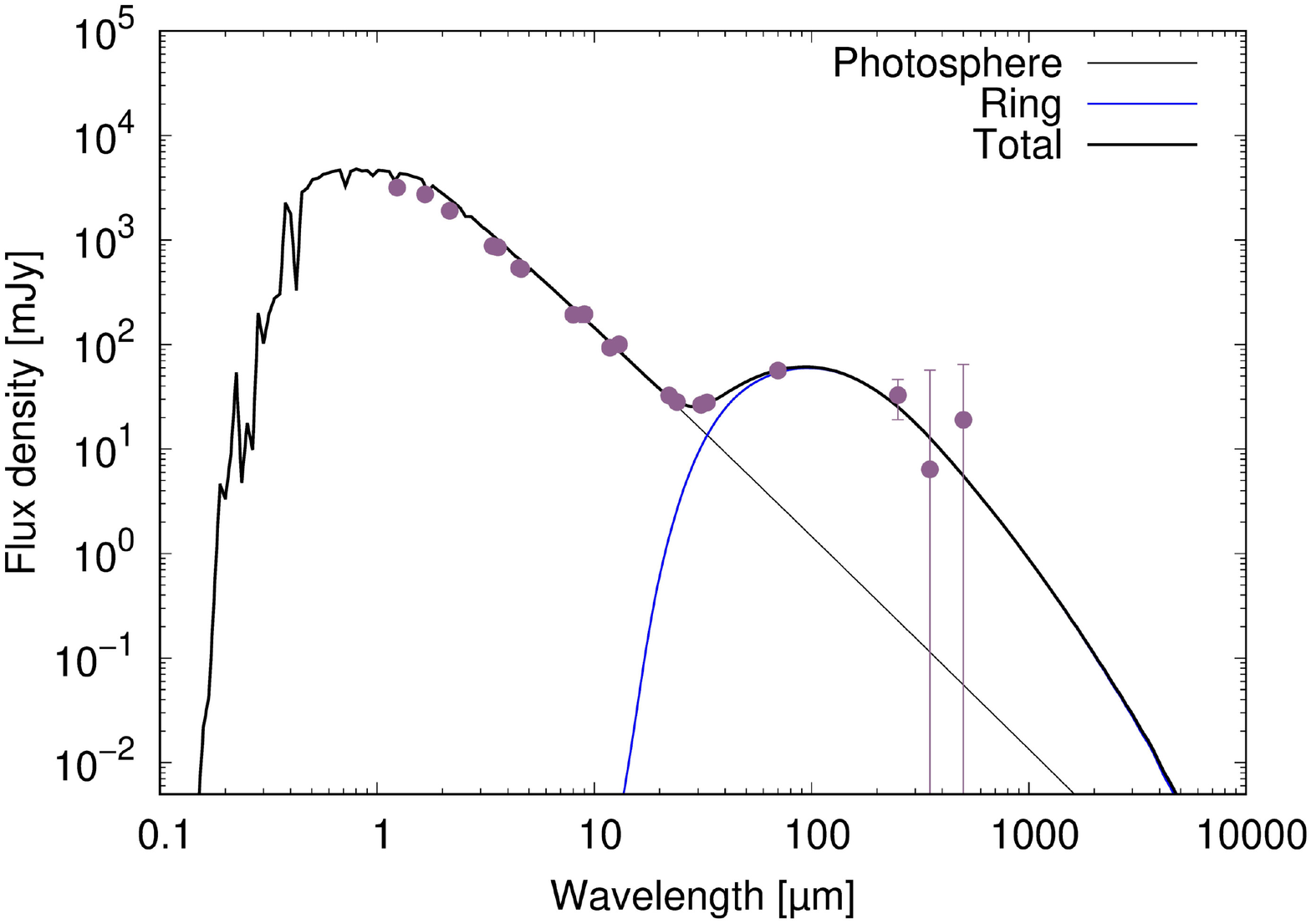}
    \caption{Flux density as a function of wavelength. The disk lies between 85 and 115~AU.}
     \label{fig:OneComponent_SED}
\end{figure}

As inferred from the scattered light image we assume the disk to lie between 85 and 115\,au. The SED is fitted using a grain size and radial distribution model with a power law of the following shape.
\begin{equation}
N(r, s) \sim s^{-q}r^{-p},
\end{equation}
where $N(r,s)drds$ is the number of grains with a size ranging from $s$ to $s+ds$ at a disk radius between $r$ and $r+dr$. The parameter $q$ is the size distribution and $p$ the radial distribution index. The fitting method is similar to the study of \cite{Pawellek2014}. 
We fix the maximum grain size to 1000\,$\muup$m and fit the minimum grain size and the size distribution index. The radial distribution index is fixed to 1.5 
resembling the profile in a small grain halo beyond a collision-dominated ring \citep{Strubbe2006}.  
We assume pure astronomical silicate \citep{Draine2003} as dust composition. 
   
\subsection{Single broad disk fit}
Given the scarcity of the photometric data in the spectral range where the disk emission dominates over the photosphere, and the relative proximity of the two belts as inferred from direct imaging, the fitting of the two components in the SED is an underdetermined problem. For the sake of simplicity, we  assume a single broad disk between 85 and 115\,au as a first approach.
The SED fit is shown in Fig.~\ref{fig:OneComponent_SED}.

We find a minimum grain size of $2.27\pm0.40\,\muup$m and a size distribution index of $3.45\pm0.22$. The reduced $\chi^2$ is found to be 11.2. This high value is caused by the relatively small error bars given in the literature. 
The corresponding mass is $1.60\times10^{-2}M_{\oplus}$.

\subsection{Double belt fit}

In the following approach, to avoid fitting an overly small number of data points with too many model parameters, we assume an inner component to be a pure blackbody ring without any size distribution and an outer component between 105 and 115\,au. We fitted the blackbody radius for the inner component, while for the outer component we fitted the minimum grain size and the size distribution index. Comparable to the SED fitting of AU~Mic \citep{Pawellek2014} the code finds the best fit by setting the mass of the inner component to zero. This means that the best fit is a single-component model, and that a pure SED fit cannot discriminate between one single disk extending from 85 to 115\,au and two rings in the same region. This is not completely unexpected, given the relatively narrow radial extent of the whole region and the relatively limited differences in terms of grain temperatures and thermal emissions.

However, our detailed SED fit rules out the need for an additional warm belt at $\sim$4\,au that was suggested by the SED fit of \citet{Chen2014} assuming pure blackbodies. This is because while a broad SED can only be explained by different spatial locations when assuming blackbodies,  a model that takes into account size distribution and size-dependent radiative properties can result in an extended range of temperatures, and thus a broader SED, for a single spatial location. We note that we do not rule out the presence of a warm belt at 4\,au, which would be undetectable with the angular resolution of SPHERE, but rather its signature in the global SED.

\section{Point sources and limit of detection}
\label{sec:contrast}
\begin{figure}[t!] 
\centering
\includegraphics[width=9cm, trim = 0.5cm 1cm 0.5cm 1cm, clip ]{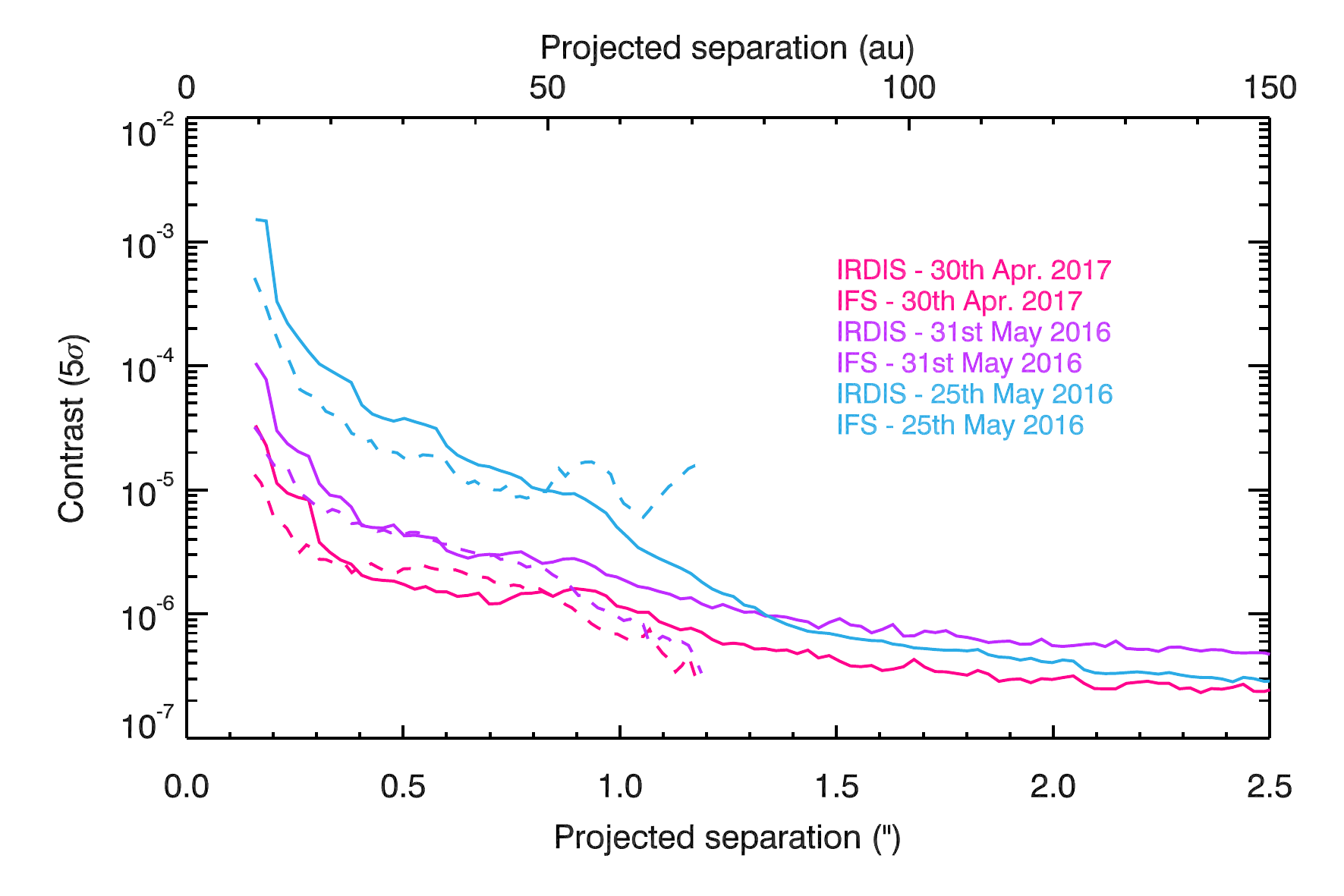}
\includegraphics[width=9cm, trim= 1.5cm 1cm 0.5cm 1cm, clip]{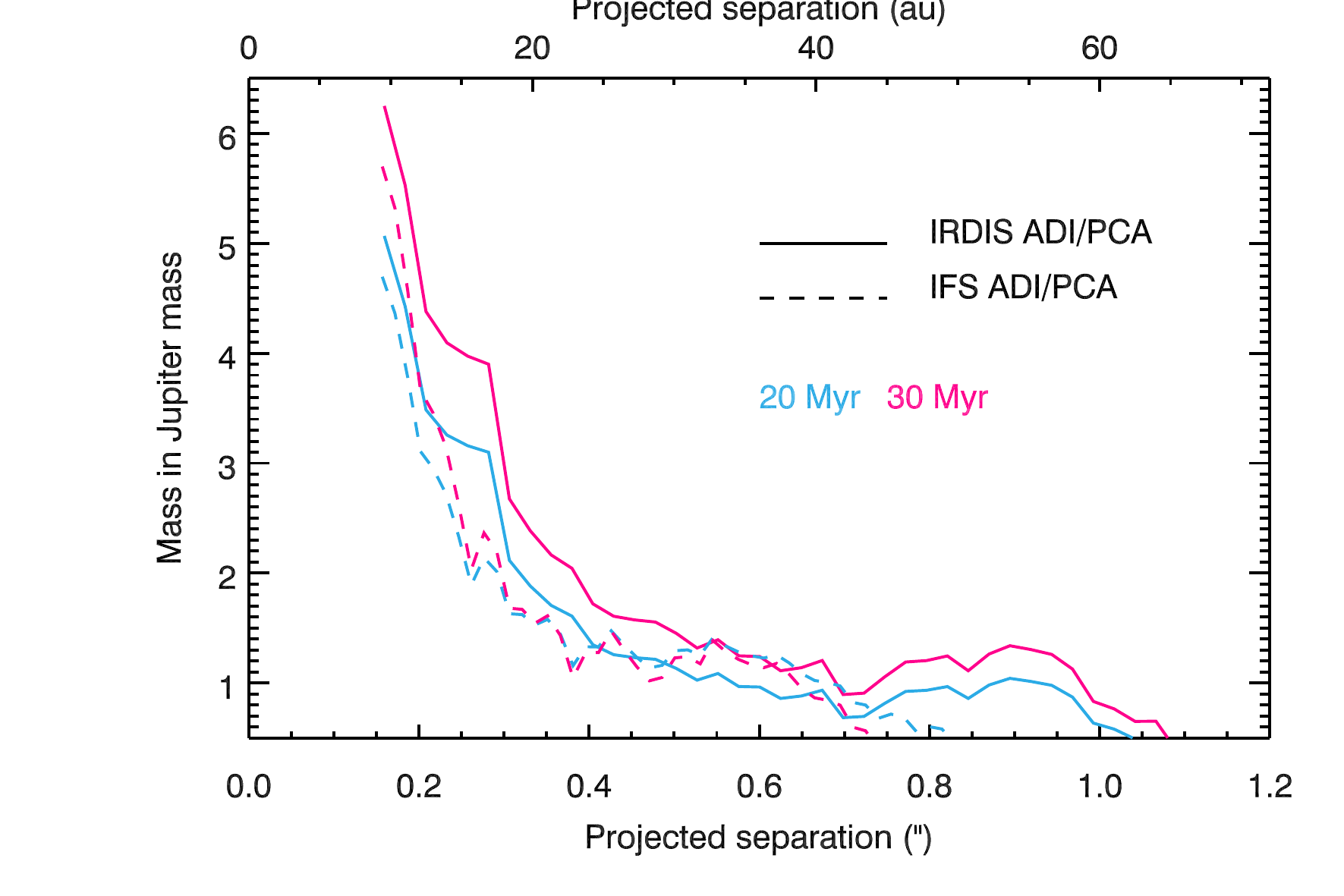}
\caption{Limits of detection in contrast (top) and converted into Jovian masses (bottom) for two age assumptions (20 and 30 Myr), using the BT-Settl atmosphere model.}
\label{fig:limdet}
\end{figure}

We identified 16 point sources in the IRDIS field of view (Fig. \ref{fig:cc}). We used \texttt{SpeCal} to determine the position of these sources in May 2016 and April 2017. A full description of astrometric errors is provided
in \citet{Galicher2018}. We used the {\it model of planet image} procedure that fits an estimation of the image of a point-source in the TLOCI reduced image. The accuracy of the fitting in the 2016
and 2017 images is 5 to 10\,mas. We accounted for the proper motion and the parallax of NZ\,Lup to predict the positions of the point-sources in the 2017 image from the measured positions in 2016 as if they were
background sources. We then compared these predictions to the measured positions in the 2017 image. None of the detected sources share the motion of NZ\,Lup. They are  consistent with background stars (see Fig. \ref{fig:ppm}). Nevertheless, some dispersions are observed between the measured and estimated positions for 2017, possibly indicating that some of these background stars also have detectable proper motions. 

The limits of the detection of point sources were estimated from the KLIP reductions both with IRDIS and IFS following the procedure described in \citet{Galicher2018}. The contrast at a particular radius is calculated from the standard deviation of the pixel contained in an annulus of 0.5$\times$FWHM in width (about 2 pixels) centered on the star.
The self subtraction inherent to ADI is estimated with fake planets injected into the data (along a spiral pattern to cover a range of separations and azimuths), and is compensated to produce the contrast plot in Fig. \ref{fig:limdet}. Inside the control radius ($\sim0.8''$ in the H band), we reached a contrast of about $10^6$ in April 2017. At a separation of 2.5$''$ the contrast is as large as $5\times10^7$. The IFS outperforms IRDIS at stellocentric distances shorter than 0.4$''$ when using ADI. In this particular case, the gain when using the spectral diversity of the IFS in addition to ADI to further improve the contrast was found to be negligible and is therefore not presented. 

These contrast values are converted into masses considering several atmosphere models calculated for the SPHERE filters, and assuming a stellar magnitude of H=6.41 for IRDIS and J=6.74 for the IFS spectral range.  The limits of detection for BT-Settl \citep{Allard2014} are shown in Fig. \ref{fig:limdet} and the two other cases, DUSTY and COND \citep{Allard2001}, are provided in the Appendix (Fig. \ref{fig:limdetDUSTYCOND}).  The models are considered unreliable for masses lower than $0.5\,M_\mathrm{Jup}$ and therefore lower values are not plotted; although the achieved contrast would in principle allow for the detection of smaller/lighter planets. A planet of $1\,M_\mathrm{Jup}$ would have been detected at a separation of 0.5$''$ (corresponding to about 30\,au in projection) according to the BT-Settl model.

\section{Discussion}
\label{sec:discussion}

Despite some difficulties to clearly distinguish between the \texttt{gap} and \texttt{2gaps} families of solutions, our analysis has established that the most likely configuration is that the NZ Lup disk extends from $\sim$85\,au to $\sim$115\,au and displays {both} a discontinuity in density (be it with a gap or a sharp transition) and a tilt in inclination between the inner and the outer parts.

 We note that HIP\,67497 is the only debris disk in which two cold but coplanar belts are observed at distances of about 60 and 130\,au \citep{Bonnefoy2017}.
The obvious reference for a disk with an inclination tilt is the beta Pictoris system seen  edge-on, with its inner $\lesssim50\,$au region tilted by a $\sim5^{\circ}$ angle with respect to the outer disk \citep[e.g.,][]{Lagrange2012}. This tilt is likely created by an outward-propagating warp induced by an off-plane inner planet \citep{Mouillet1997, Augereau2001}, later identified as being the imaged $\beta$\,Pic\,b planet \citep{Lagrange2010, Lagrange2012}. 
HST images of the beta Pic disk \citet{Golimowski2006} give the impression of two separate disks due to the fortuitous  alignment of the line of nodes and the line of sight. The NZ Lup system could have a morphology similar to $\beta$ Pictoris, but with the line of nodes nearly perpendicular to the line of sight. The mutual inclinations of the belts are of the same order, $\sim4-5^{\circ}$.
However, in the present case, we believe that a single undetected off-plane planet cannot account for all the observed characteristics of the disk.
A planet located between the inner and outer parts of the disk (at around $90-100$\,au) would for example carve out a gap or a density drop \citep[e.g.,][]{Lazzoni2018} 
and launch both inward- and outward-propagating waves that would warp the disk inside and outside its orbit. However, we would expect the inner and outer warp to propagate at roughly the same speed, which would tend to align both of them towards the same plane. One way to alleviate this problem would be for the potential perturbing planet to be placed inside the inner edge of the disk ($\leqslant85$\,au). In this case, the warp would propagate outwards and would first reach the inner regions of the disk before affecting its outer parts, thus creating a de facto tilt between these two regions. However, such a planet would not be able to create a density gap or discontinuity in the middle of the disk. 

An additional characteristic of the NZ Lup disk that the planet-induced warp scenario cannot explain is its very small vertical thickness, with $h \lesssim0.02$. This is at odds with the theoretical prediction that the warped regions should be puffed up to an opening angle roughly equal to twice the inclination of the planet \citep{Nesvold2015}, meaning that this opening angle should be comparable to the off-plane tilt of the warp. Indeed, in the present case, this would lead to an opening angle of $\sim5^{\circ}$, which is at least a factor four higher than the value constrained by our parametric fitting.

Our modeling results suggest that the region inwards of 80\,au is cleared of dust. If one assumes that planetesimals exist at all radii in the initial stage of planet formation and that gaps or cavities in debris disks are created by the emerging planets, one can assess the number of planets and their minimum masses necessary to create such a gap between 10 and 80\,au during the lifetime of the system (i.e., 16 Myr). To do that, we use the numerical work of \citet{Shannon2016} to constrain the lower limit in mass for planets to have depleted the inner regions. To clear a gap between 10 and 80\,au over 16 Myr, we find that two planets of $> 0.65\,M_\mathrm{Jup}$ are needed. \citet{Shannon2016} assume equal-mass planets separated by 20 mutual Hill radii (i.e., the typical separation between planets in Kepler multi-planet systems). Using the planet mass upper limits from this paper together with the lower limit that we have just derived, we find that the potential planet able to warp the inner cold disc would have a mass of between 0.65 and $2\,M_\mathrm{Jup}$.

\citet{Lee2016}  investigated a variety of debris disk morphologies considering only a few parameters, the viewing angles, and a planet eccentricity orbiting inside a parent-belt of planetesimals. In particular, their model is able to produce a double wing geometry (also referred to as a moth geometry), if  the planet eccentricity is large enough ($\sim$0.7), the apoastron of dust orbits is towards the observer, and the system is nearly edge-on (5-10$\degb$). Depending on whether the tail of dust particles is directed towards or away from the observer, the double wing geometry can morph into a bar geometry. This picture can describe rather well the case of HD\,32297 where a faint bar with an offset from the main ring is seen from the HST STIS image \citep{Schneider2014}. 
However, the bar is nearly parallel to the midplane, while the two wings deviate from each other with the stellocentric distance. These behaviors do not match the image of NZ\,Lup.
The disk spine in Fig.  \ref{fig:spine} shows that the second component is converging towards the midplane, which compounds  the hypothesis of two separated rings.

Some multiple belts observed in a few debris disks (HD\,131835 or HD\,141569) are also suspected to be the result of the photoelectric effect generated by a gas--disk interaction \citep{Lyra2013}, although inclinations of such belts are expected to be uniform. 
However, no gas detection has been reported so far for NZ Lup and therefore a planet scenario would be more likely. In any case, the two aforementioned debris disks feature several thin belts as expected for scenarios involving gas, while NZ Lup has only two. 
It is therefore not easy to find a straightforward explanation for the peculiar characteristics of this system. More sophisticated dynamical and numerical explorations should be carried out in the future in order to address this issue as well as a systematic search for planets.

\section{Summary}

Here we summarize the results of our analysis on the NZ\,Lup debris disk. 

\begin{itemize}
\item[$\bullet$]{The angular resolution and contrast achieved with SPHERE reveal a globally very inclined debris disk ($\sim$85\deg) with a disk splitting attributed to the presence of two noncoplanar planetesimal belts. 
The sizes of these belts are derived from modeling with GRaTer. Assuming some variations of the density between the belts, we determine the plausible range of stellocentric distances: 80-95\,au for the inner belt, and 95-120\,au for the outer belt.}
\item[$\bullet$]{The modeling favors a configuration where the outer belt is more inclined than the inner belt by $\sim$5\deg. The reversed geometry (inner belt is more inclined) is significantly worse in terms of matching the data and is therefore ruled out.  }
\item[$\bullet$]{The relative intensity between these two belts necessarily implies a variation of density with a discontinuity, which naturally generates a gap in scattered light. Models with a true dust depletion (or density gap) between the belts more closely match  the images, although the difference is marginal with other model families where the region in between the belts is not completely depleted.}
\item[$\bullet$]{The fine structure revealed by direct imaging is not measurable in the SED. However, the scarcity of photometric data in the far-IR supports the argument for a broad component, which is compatible with the position of the belts inferred from direct imaging. A closer warm dust component as proposed from a former analysis of the SED cannot be confirmed either from the reanalysis of the SED or from SPHERE observations.}\item[$\bullet$]{No co-moving candidates could be identified in the SPHERE data for all three available epochs, but the limit of detection reaches a lower limit of $1\,M_\mathrm{Jup}$ at a separation of 0.6$''$ (projected separation of $\sim$4\,au).}
\item[$\bullet$]{The scattered light of the disk is also identified for the first time in the visible using HST archival data. Some characteristics can be recovered as compared to near-IR imaging with SPHERE but the data lacks contrast at short angular separations to confirm the presence of the double belt.}
\item[$\bullet$]{Explaining the distribution of the dust in the form of two belts with a single planet has some theoretical shortcomings. Multiple planets could be required but their masses would be much lower than the limit of detection. }
\end{itemize}

The debris disk around the G-type star NZ\,Lup features a rare case of multiple belts observed in a planetary system. Only four such systems have been identified with direct imaging, at least two of which are gas-rich. Therefore, this system is crucial in the context of understanding the last stages of planetary system evolution. It should be a prime target for future high-contrast facilities.  The identification of the double belt pattern in SPHERE data was made difficult due to the high inclination of the disk and the ADI-induced artifacts. We are planning a follow-up investigation of this target at shorter wavelengths, with ZIMPOL for instance, to take advantage of a higher angular resolution, as well as in polarimetry (both near-IR and visible) to provide diversity in the phase function.

\bibliography{35135corr}

\begin{acknowledgement}
SPHERE is an instrument designed and built by a consortium consisting of IPAG (Grenoble, France), MPIA (Heidelberg, Germany), LAM (Marseille, France), LESIA (Paris, France), Laboratoire Lagrange
(Nice, France), INAF–Osservatorio di Padova (Italy), Observatoire de Genève (Switzerland), ETH Zurich (Switzerland), NOVA (Netherlands), ONERA (France) and ASTRON (Netherlands) in collaboration with ESO. SPHERE was funded by ESO, with additional contributions from CNRS (France), MPIA (Germany), INAF (Italy), FINES (Switzerland) and NOVA (Netherlands).  SPHERE also received funding from the European Commission Sixth and Seventh Framework Programmes as part of the Optical Infrared Coordination Network for Astronomy (OPTICON) under grant number RII3-Ct-2004-001566 for FP6 (2004–2008), grant number 226604 for FP7 (2009–2012) and grant number 312430 for FP7
(2013–2016). 
French co-authors also acknowledge financial support from the Programme National de Planétologie (PNP) and the Programme National de Physique Stellaire (PNPS) of CNRS-INSU in France. 
This work has also been supported by a grant from the French Labex OSUG@2020 (Investissements d’avenir – ANR10 LABX56). 
The project is supported by CNRS, by the Agence Nationale de la Recherche (ANR-14-CE33-0018). 
Italian co-authors acknowledge support from the "Progetti Premiali" funding scheme of the
Italian  Ministry  of  Education,  University,  and  Research.
This work has been supported by the project PRIN-INAF 2016 The Cradle of Life - GENESIS-
SKA (General Conditions in Early Planetary Systems for the rise of life with SKA). 
C.\,P. acknowledges support from the ICM (Iniciativa Cient\'ifica Milenio) via the N\'ucleo Milenio de Formaci\'on Planetaria grant. 
Finally, this work has made use of the the SPHERE Data Centre, jointly operated by OSUG/IPAG (Grenoble), PYTHEAS/LAM/CESAM (Marseille), OCA/Lagrange (Nice) and Observatoire de Paris/LESIA (Paris). We thank P. Delorme and E. Lagadec (SPHERE Data Centre) for their efficient help during the data reduction process. 

\end{acknowledgement}

\clearpage

\begin{appendix}

\section{Signal-to-noise-ratio maps}
\begin{figure*}[h] 
\centering
\includegraphics[width=18cm]{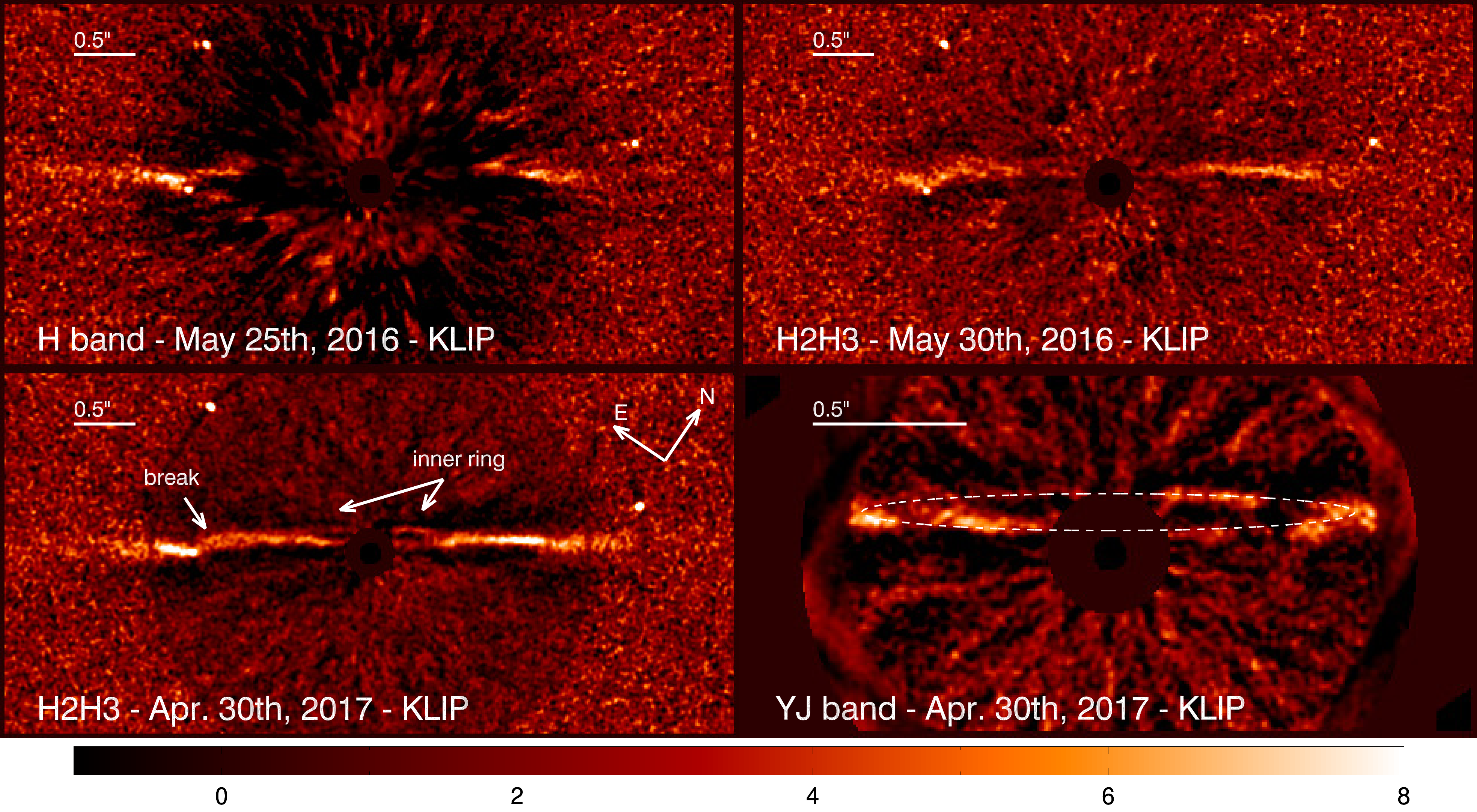}
\caption{Signal-to-noise-ratio map per resolution element (indicated in the color bar). The disk midplane is aligned with the horizontal direction and the displayed field of view is $6''\times3''$ for IRDIS and $2.4''\times1.2''$ for the IFS.}
\label{fig:snr}
\end{figure*}

\clearpage
\section{Frequencies of model parameters}
\begin{figure}[h] 
\centering
\includegraphics[width=9cm]{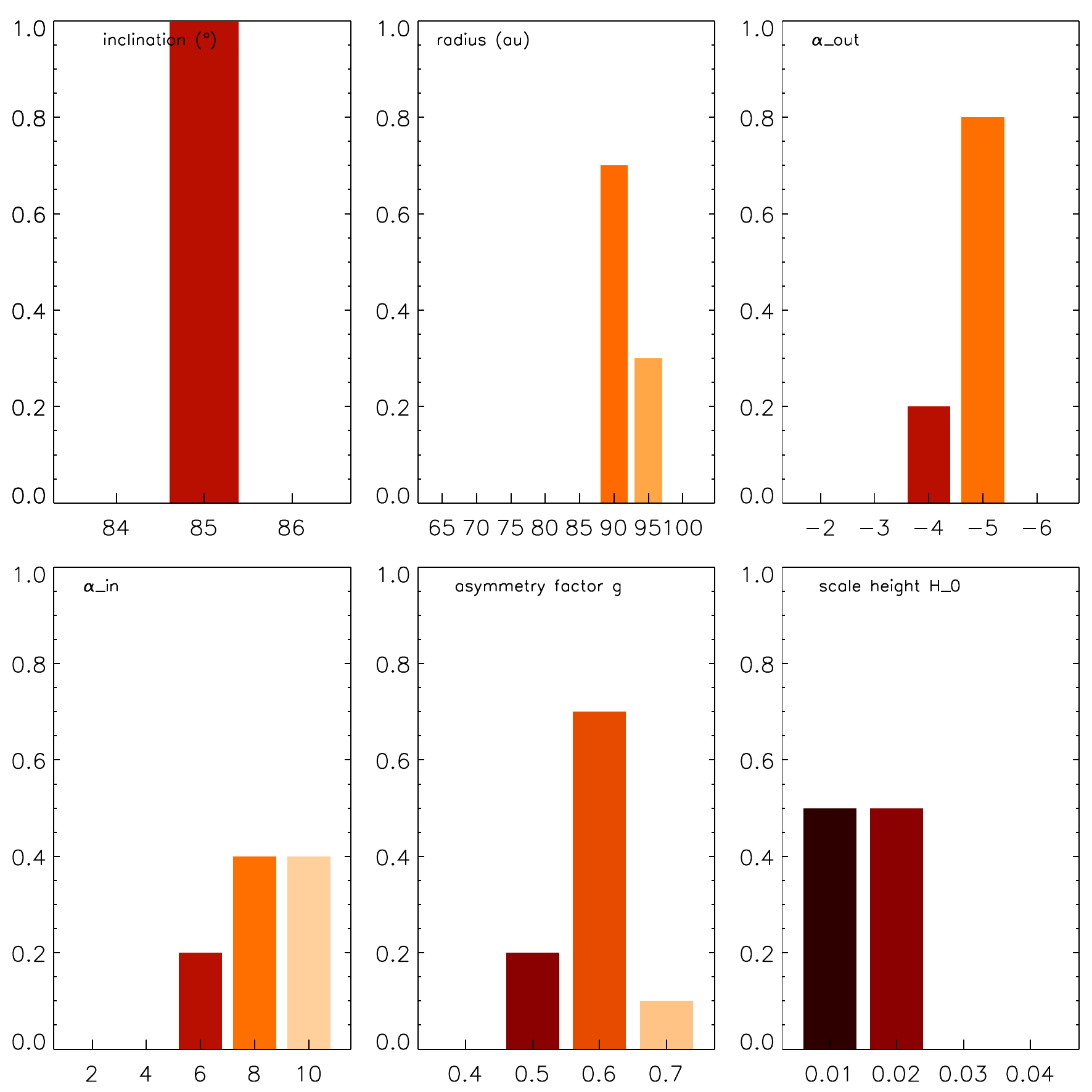}
\caption{Frequencies of parameters obtained for the 1\% best models in the single-belt \texttt{mask1} case.}
\label{fig:freq_1belt}
\end{figure}
\begin{figure}[h] 
\centering
\includegraphics[width=9cm]{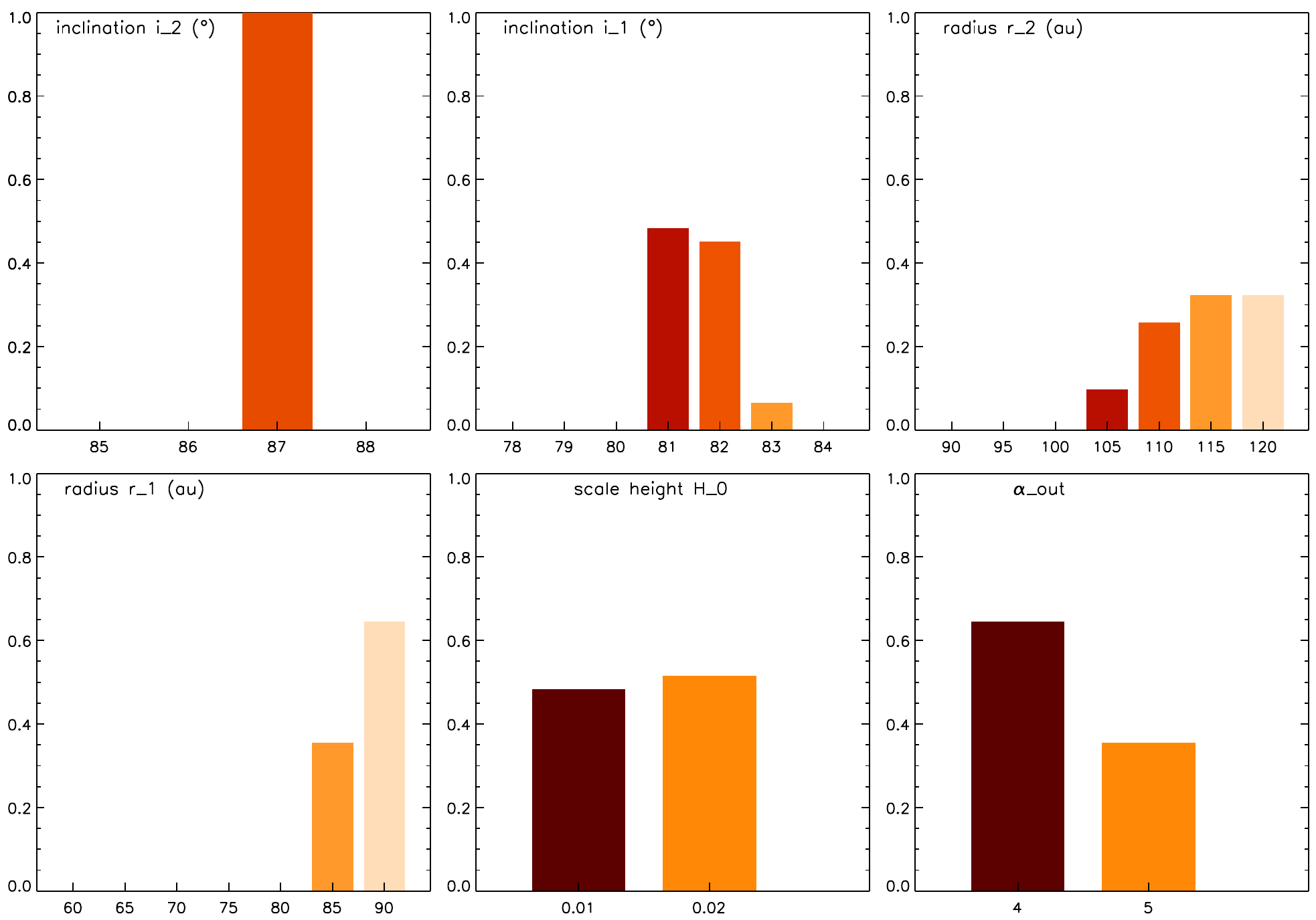}
\caption{Frequencies of parameters obtained for the 1\% best models in the two-belt \texttt{gap} case.}
\label{fig:freq_truegap}
\end{figure}
\begin{figure}[h] 
\centering
\includegraphics[width=9cm]{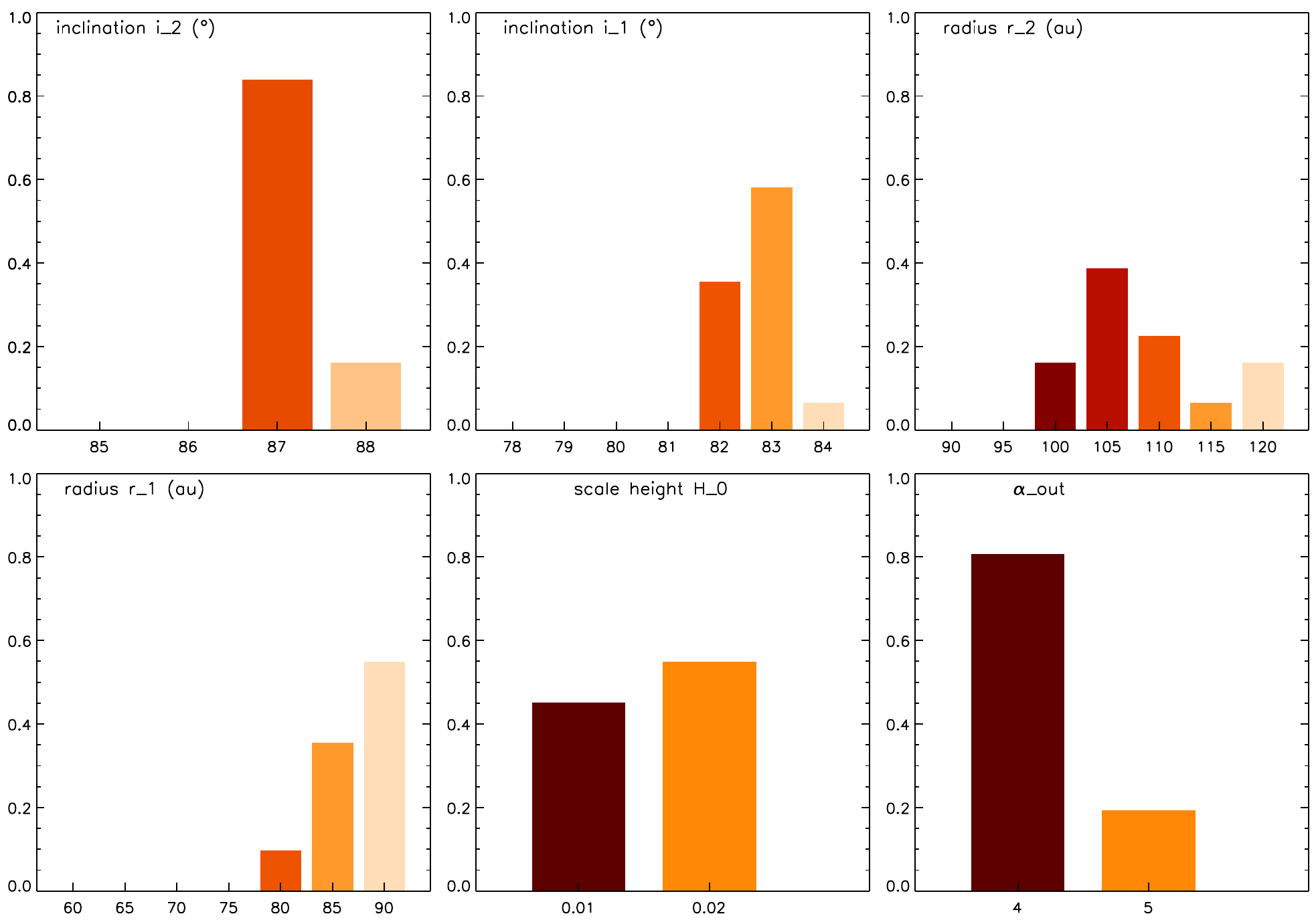}
\caption{Frequencies of parameters obtained for the 1\% best models in the two-belt \texttt{fill} case.}
\label{fig:freq_truefill}
\end{figure}
\begin{figure}[h] 
\centering
\includegraphics[width=9cm]{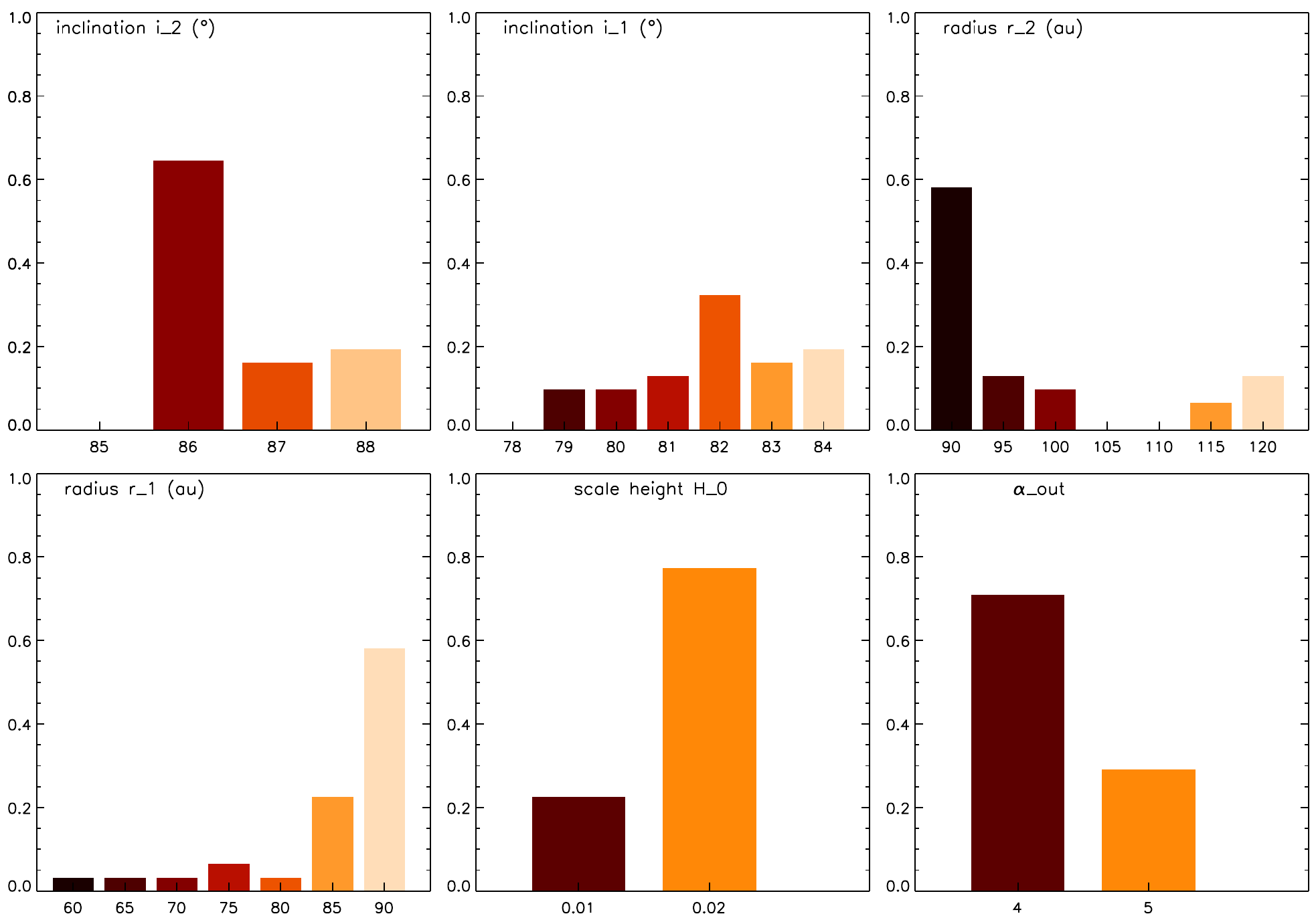}
\caption{Frequencies of parameters obtained for the 1\% best models in the two-belt \texttt{ring} case.}
\label{fig:freq_justaring}
\end{figure}

\clearpage
\section{Inclination effect on the second belt}
\begin{figure}[h] 
\centering
\includegraphics[width=9cm]{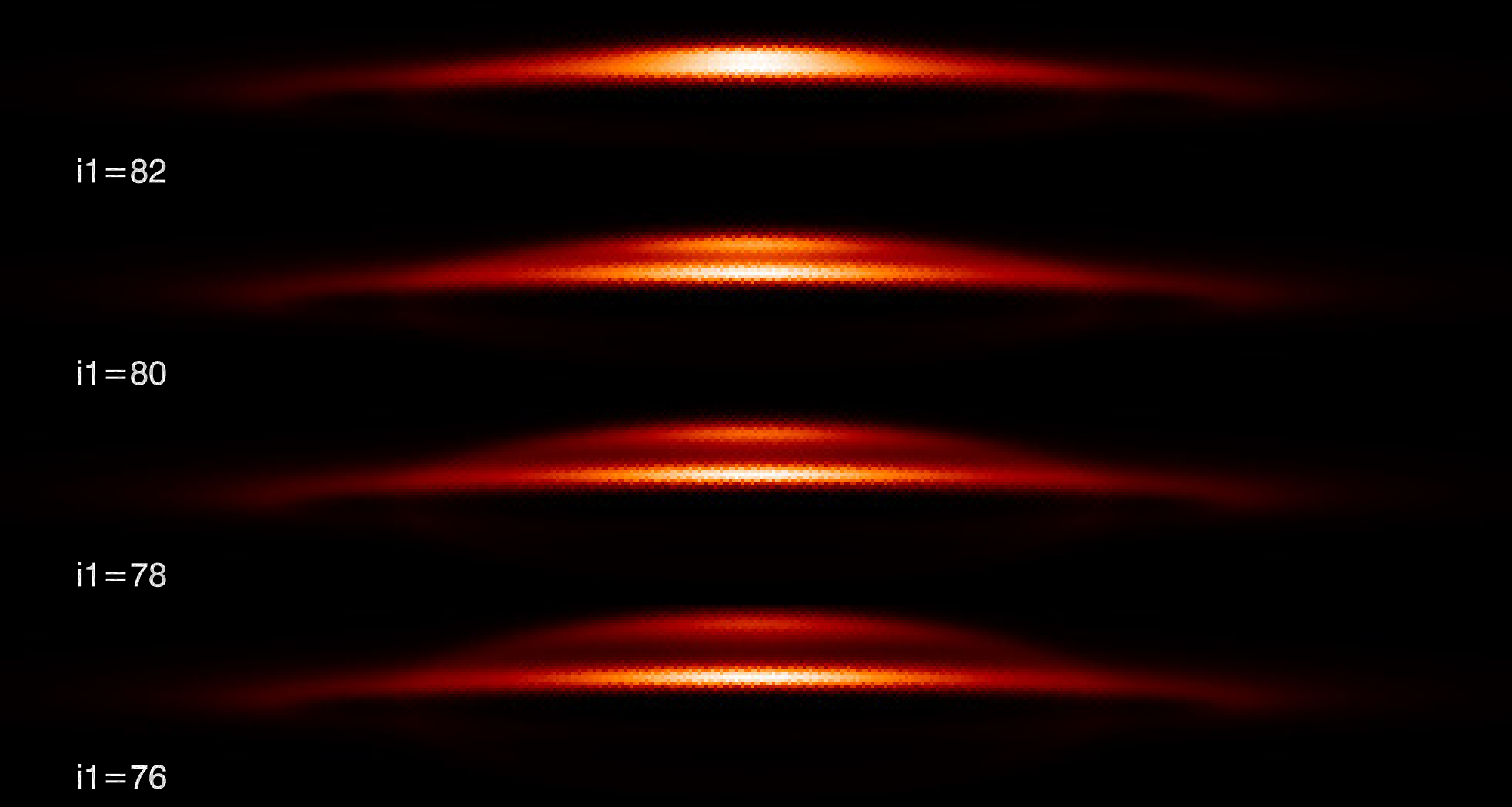}
\caption{Two-belt models assuming the outer component is inclined by 87\deg     while the inner belt is (from top to bottom) inclined by 82, 80, 78 and 76\deg. The relative intensity evolves according to the mutual inclination. }
\label{fig:inclinationeffect}
\end{figure}

\clearpage

\section{Comparison of data and best model images}

\begin{figure}[h] 
\centering
\includegraphics[width=9cm]{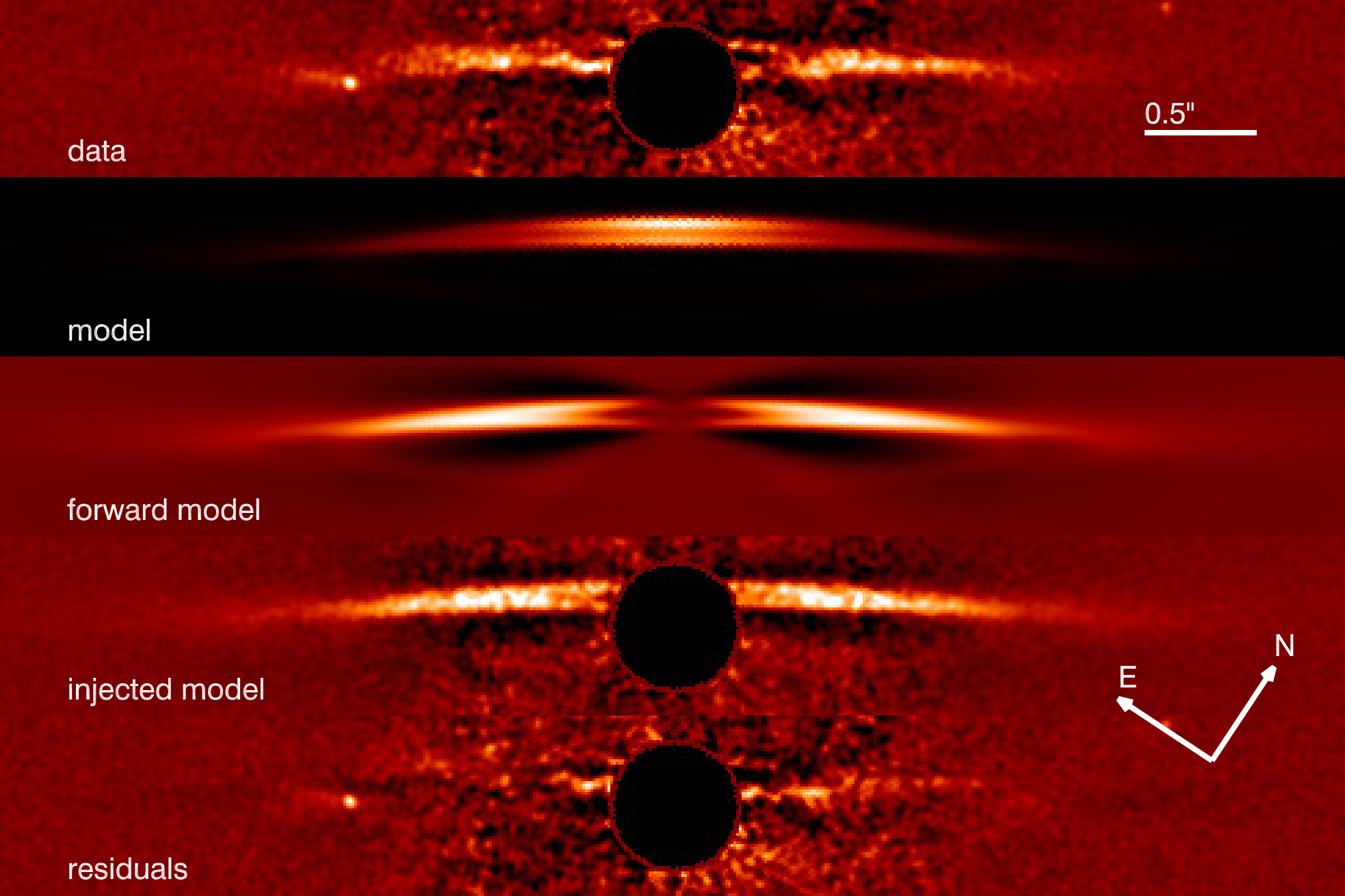}
\caption{
From top to bottom: Original data from April 2017 (same as in Fig. \ref{fig:images}), 
the raw model before PSF convolution, the corresponding forward model, and the best model injected in the data, and the residuals (\texttt{gap} case: $i_1=82\degb$,  $i_2=87\degb$, $r_1=90$\,au,  $r_2=115$\,au).  The field of view is $6''\times0.8''$.}
\label{fig:datavsmodels_truegap}
\end{figure}
\begin{figure}[h] 
\centering
\includegraphics[width=9cm]{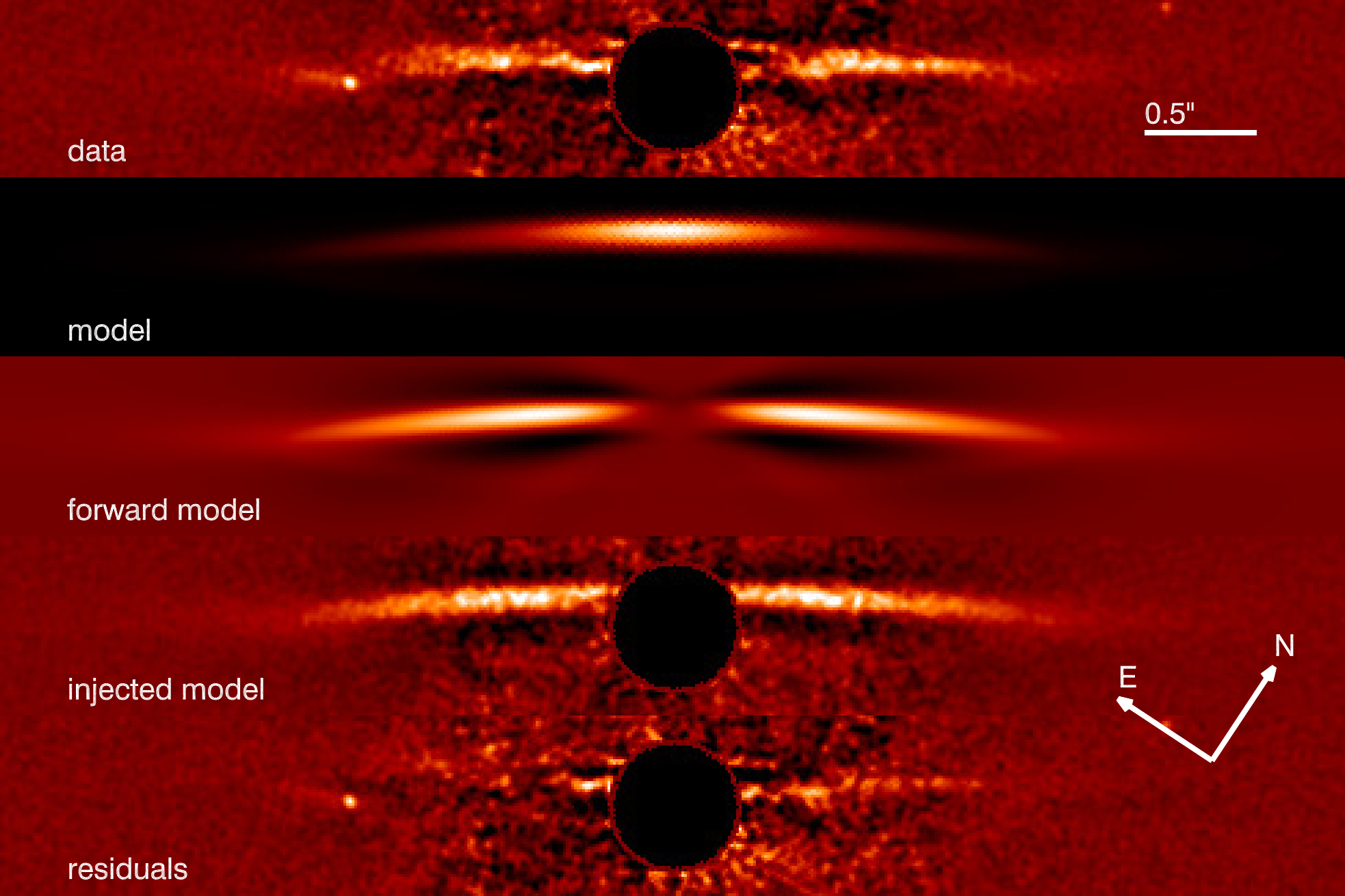}
\caption{From top to bottom: Original data from April 2017 (same as in Fig. \ref{fig:images}), 
the raw model before PSF convolution, the corresponding forward model, and the best model injected in the data, and the residuals (\texttt{fill} case: $i_1=83\degb$,  $i_2=87\degb$, $r_1=90$\,au,  $r_2=105$\,au).  The field of view is $6''\times0.8''$.}
\label{fig:datavsmodels_truefill}
\end{figure}

\begin{figure}[h] 
\centering
\includegraphics[width=9cm]{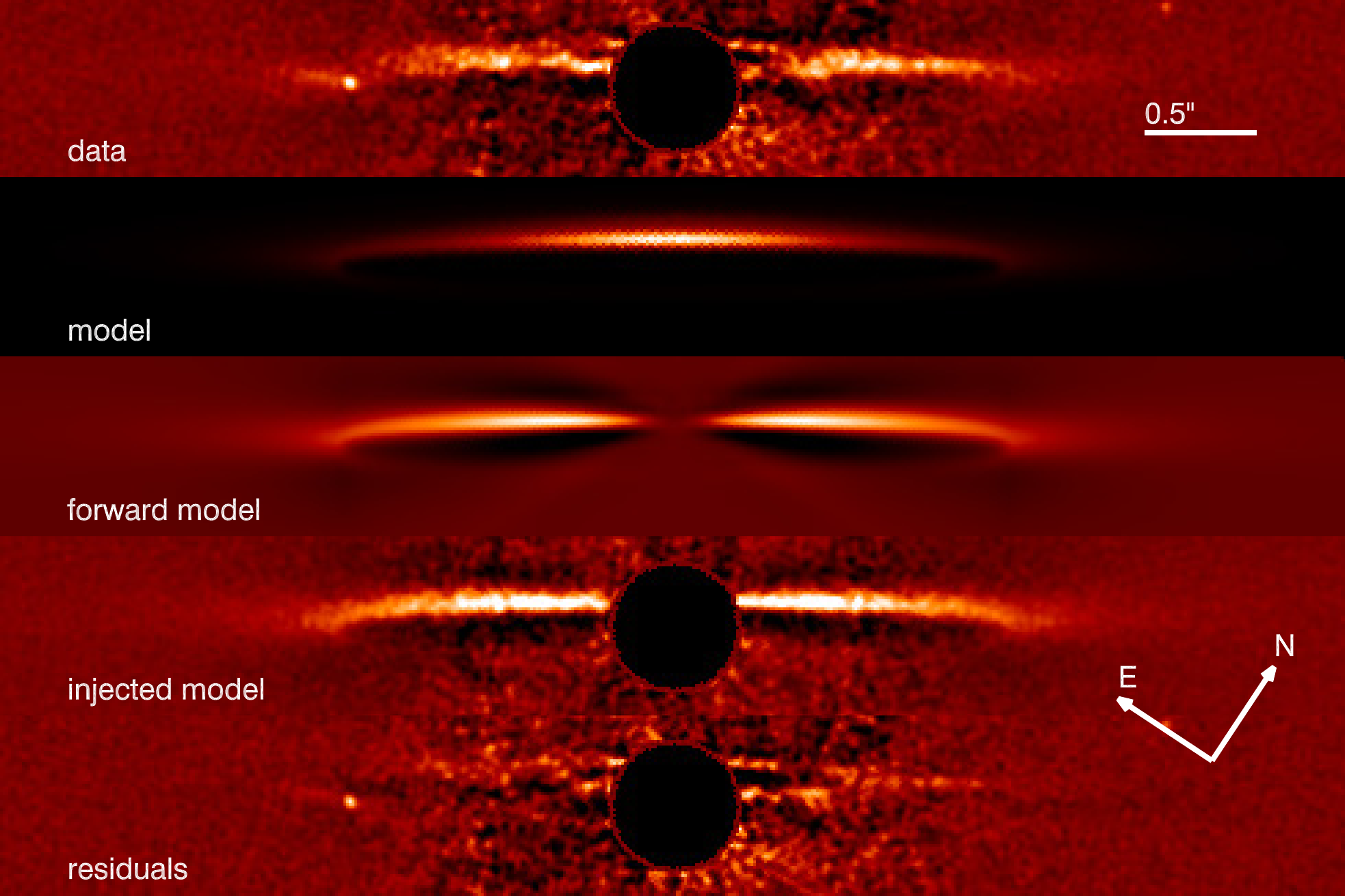}
\caption{From top to bottom: Original data from April 2017 (same as in Fig. \ref{fig:images}), 
the raw model before PSF convolution, the best model injected in the data, and the residuals (\texttt{ring} case: $i_1=86\degb$,  $i_2=82\degb$, $r_1=900$\,au,  $r_2=90$\,au).  The field of view is $6''\times0.8''$.}
\label{fig:datavsmodels_justaring}
\end{figure}

\clearpage
\section{Characterization of point sources in the IRDIS field of view}
\begin{figure}[h]
\centering
\includegraphics[width=10cm]{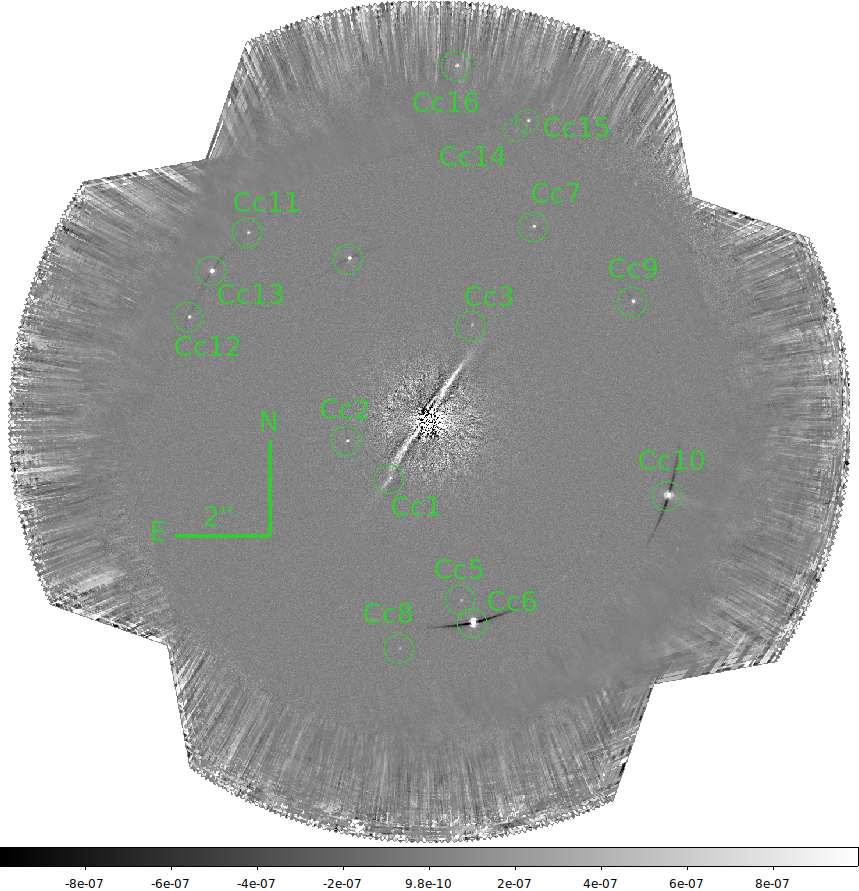}
\caption{Positions of all identified points sources in the IRDIS field of view.}
\label{fig:cc}
\end{figure}
\begin{figure*}[h]
\centering
\includegraphics[width=15cm]{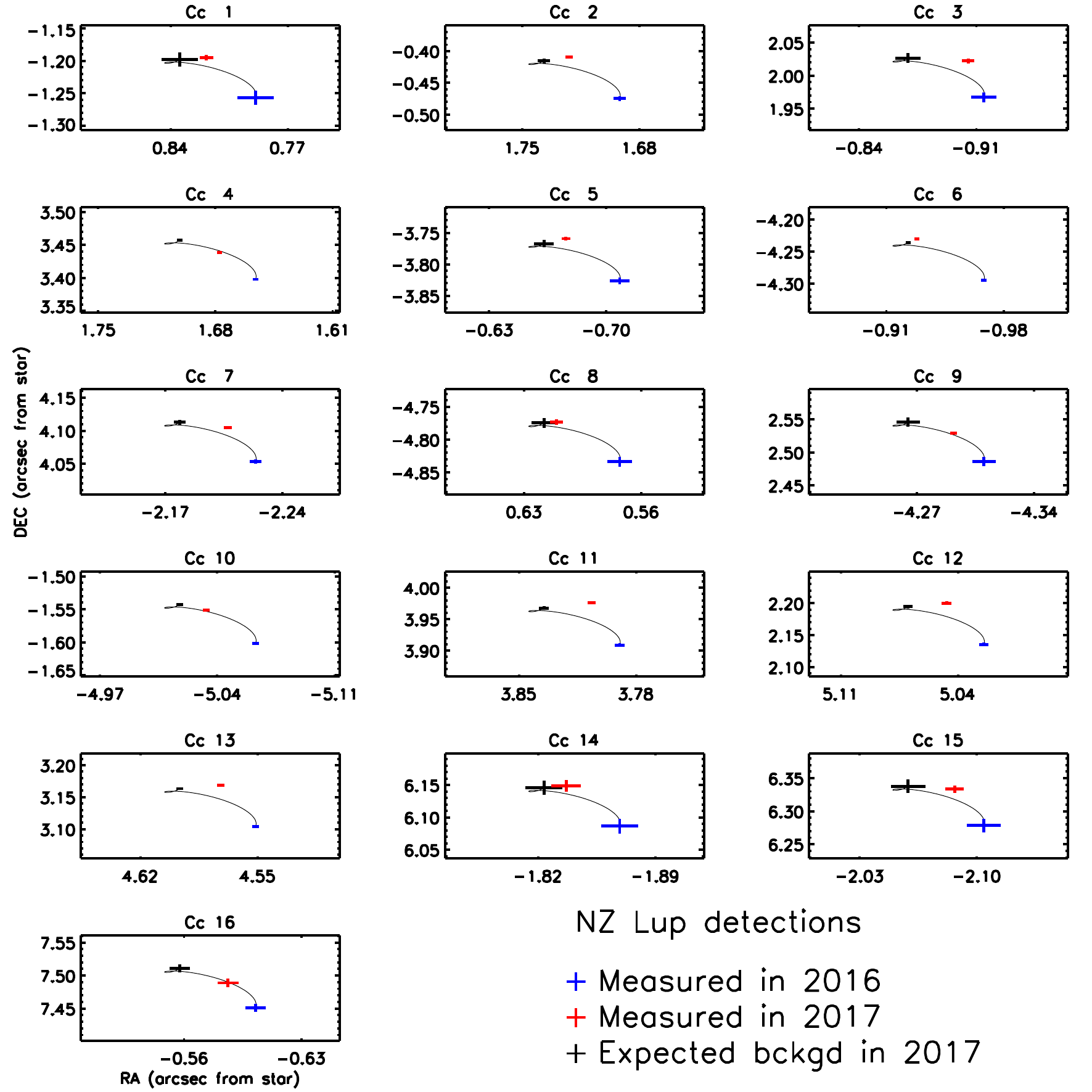}
\caption{Positions of all points sources in the IRDIS field of view as measured at two epochs (2016-05-31 and 2017-04-30) compared to the expected positions if they were background stars.}
\label{fig:ppm}
\end{figure*}

\clearpage
\section{Limits of detection for DUSTY and COND atmosphere models}
\begin{figure*}[h]
\centering
\includegraphics[width=9cm]{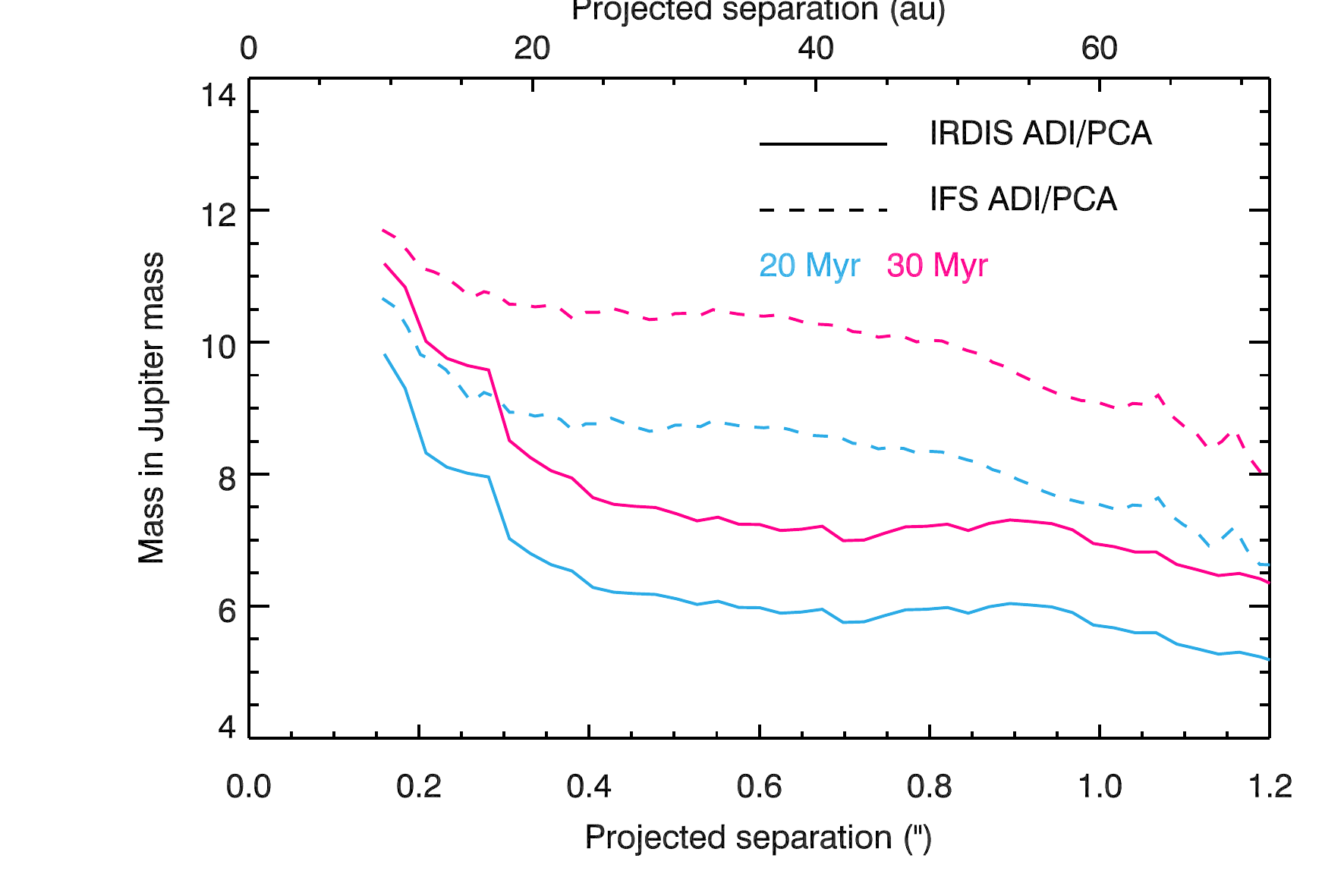}
\includegraphics[width=9cm]{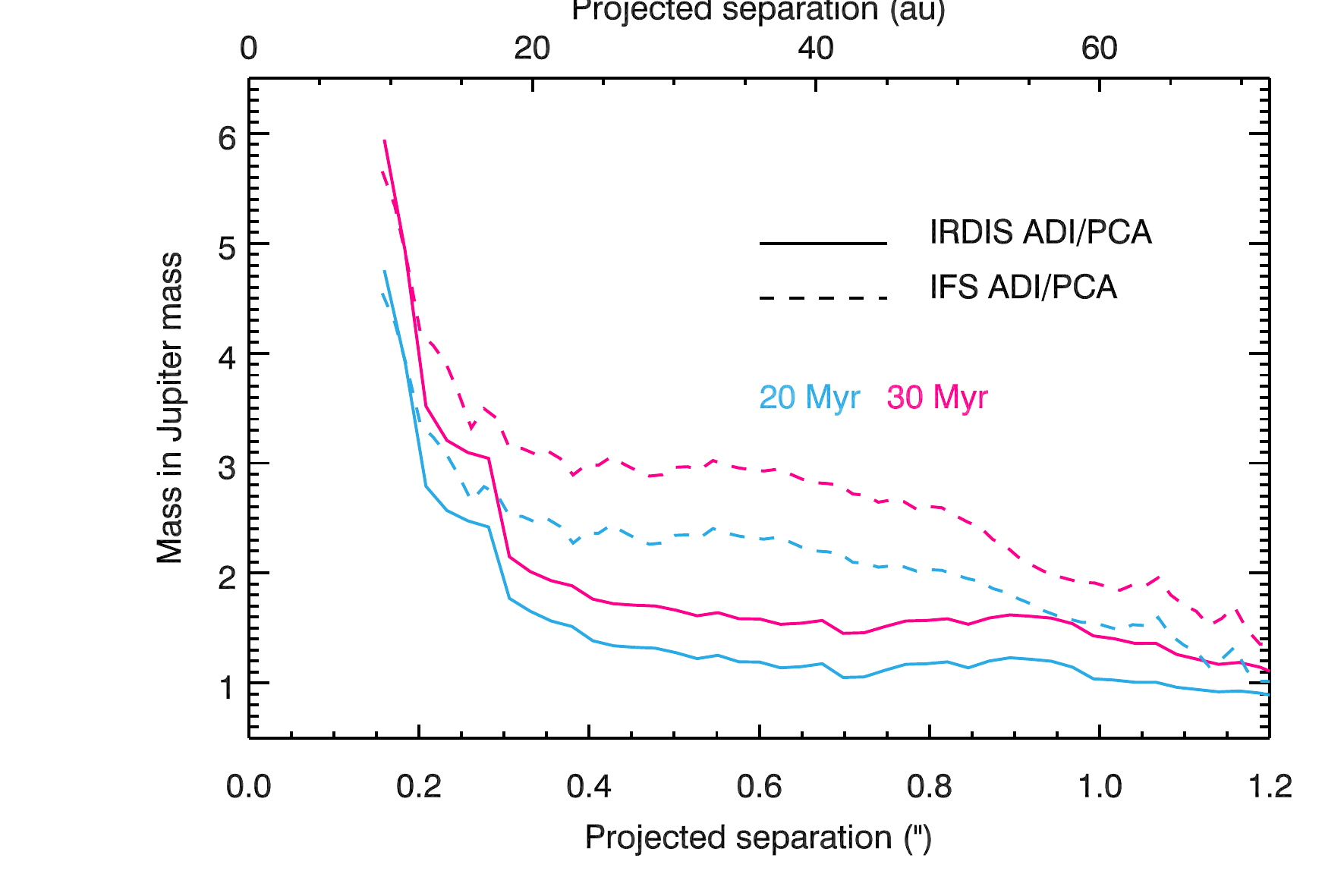}\caption{As in Fig. \ref{fig:limdet} but for the DUSTY (left) and COND (right) model.}
\label{fig:limdetDUSTYCOND}
\end{figure*}

\end{appendix}

\end{document}